\documentclass[10pt]{article}
\usepackage[top=75pt,bottom=95pt,left=45pt,right=45pt]{geometry}

\usepackage{amsmath,amsfonts,amssymb}
\usepackage{subfig,placeins}
\usepackage{chemarr,feynmp,color}
\usepackage[pdftex]{graphicx}

\usepackage[mathscr]{eucal}
\usepackage{enumerate}

\usepackage{titlesec}
\definecolor{MSBlue}{rgb}{0.0, 0.0, 0.3}
\titleformat*{\section}{\Large\bfseries\sffamily\color{MSBlue}}
\titleformat*{\subsection}{\large\bfseries\sffamily\color{MSBlue}}
\titleformat*{\subsubsection}{\normalsize\bfseries\sffamily\color{MSBlue}}

\renewcommand\vec[1]{\boldsymbol{#1}} 
\renewcommand\matrix[1]{\,\mathrm{\underline{#1}}\,} 
\def\dd{\text{d}}

\def\hc{\text{h.c.}}

\newenvironment{dTwoPoint}
 {\begin{fmfgraph*}(80,30)
 \fmfleft{i}\fmfright{o}
 }
 {\end{fmfgraph*}}

\renewcommand{\fmfct}[1]{\fmfv{decor.shape=circle,decor.filled=0,decor.size=2thick}{#1}}

\begin{document}

\title{\bfseries\sffamily\color{MSBlue}System size expansion using Feynman rules and diagrams}


\author{\large\sffamily\color{MSBlue} Philipp Thomas\footnote{School of Mathematics and Maxwell Institute for Mathematical Sciences, University of Edinburgh, Edinburgh EH9 3JZ, United Kingdom} \footnote{School of Biological Sciences, University of Edinburgh, Edinburgh EH9 3JH, United Kingdom}, Christian Fleck\footnote{Laboratory for Systems and Synthetic Biology, Wageningen University, Netherlands}, Ramon Grima$^\dagger$, Nikola Popovi\'c$^{\ast}$ }

\maketitle

\abstract{
Few analytical methods exist for quantitative studies of large fluctuations in stochastic systems. In this article, we develop a simple diagrammatic approach to the Chemical Master Equation that allows us to calculate multi-time correlation functions which are accurate to a any desired order in van Kampen's system size expansion. Specifically, we present a set of \emph{Feynman rules} from which this diagrammatic perturbation expansion can be constructed algorithmically. {We then apply the methodology to derive in closed form} the leading order corrections to the linear noise approximation of the intrinsic noise power spectrum for general biochemical reaction networks. Finally, we illustrate our results by describing noise-induced oscillations in the Brusselator reaction scheme which are not captured by the common linear noise approximation.
}

\section{Introduction}

The quantification of intrinsic fluctuations in biochemical networks is becoming increasingly important for the study of living systems, as some of the key molecular players are expressed in low numbers of molecules per cell \cite{mcadams1997}. The most commonly employed method for investigating this type of cell-to-cell variability is the stochastic simulation algorithm (SSA) \cite{gillespie2007}. While {the SSA} allows for the exact sampling of trajectories of discrete intracellular biochemistry, it is {computationally expensive, which} often prevents one from obtaining statistics for large ranges of the biological parameter space. A {widely used} approach to overcome these limitations has been the linear noise approximation (LNA) \cite{ElfEhrenberg,hayot2004,paulsson2005}.

The LNA is obtained to leading order in van Kampen's system size expansion \cite{vanKampen1976}. Within this approximation, the average concentrations are given by the solution of the macroscopic rate equations, while the fluctuations about {that solution} are Gaussian. It is however the case that the LNA can in some instances be in severe disagreement with the solution of the corresponding Chemical Master Equation (CME) which it approximates. {One such} scenario is encountered in the study of networks that involve bimolecular reactions in which some species are present at low molecule numbers. The system size expansion hence offers the possibility to investigate these discrepancies systematically by taking into account the next order terms that go beyond the LNA. In the past, these higher order terms have been successfully employed in demonstrating deviations of the mean concentrations in stochastic biochemical networks from traditional rate equation models \cite{grima2010,ramaswamy2012}, in elucidating the dependence of the coefficients of variation on biochemical network parameters \cite{BMC}, and in estimating two-time correlation functions numerically \cite{scott2012}. Different approaches have also been applied to obtain related estimates analytically in particular cases \cite{chaturvedi1978,thomas2013}. In practice, the investigation of these effects is important for inferring biological network parameters and for predicting biochemical dynamics from {mathematical} models.

A common obstacle for {the requisite} analysis on the basis of the system size expansion is the increasing algebraic complexity of the resulting equations, as is commonly encountered with higher order perturbation theories. A common remedy is to replace the CME by a truncated partial differential equation that is accurate to some given order in powers of the inverse square root of the system size \cite{vanKampen1976,grima2011,grima2011const}. For biochemical systems, the latter is given by the reaction volume. In practice, however, the calculation of higher order terms can become cumbersome because of the difficulty {involved in} solving these differential equations with increasing numbers of terms and relations. In physics, in particular in the theory of elementary particles or statistical fields, powerful techniques that overcome this complexity are Feynman diagrams and the associated rules from which these are constructed \cite{srednicki}.

{While the construction of path-integrals for {master equations} is common practice for instance via the Doi-Peliti formalism or similar techniques \cite{peliti1985,droz1994,kamenev2011,tauber2014}, we build on the functional integral representation introduced by Tirapegui and Calisto \cite{calisto1993}. The particular advantage of this formulation is that it {enables us} to construct a path-integral for the molecular concentrations and, hence, {that it} facilitates a straightforward system size expansion. {Here, we} develop the system size expansion of the CME using Feynman diagrams, which} allows us to calculate arbitrary multi-time correlation functions to arbitrary order in the (inverse) system size. To that end, we express the expansion of {these} functions to each order as a finite sum of diagrams. Specifically, we provide a set of \emph{Feynman rules} from which the system size expansion can be constructed algorithmically, or simply ``drawn'' in {diagrammatic form}. Each of these diagrams is naturally associated with a certain term in the expansion of the correlation functions, and can be evaluated using a simple dictionary. {Our} methodology hence presents an efficient bookkeeping device for the construction of closed-form expressions from the system size expansion without the need {for} solving a large number of coupled ordinary differential equations{. Its} development is the principal aim of the present manuscript.

The methodology is subsequently applied to determine the finite volume corrections to the LNA of the intrinsic noise power spectrum in general biochemical reaction networks. The latter is defined by the Fourier transform of the autocorrelation function. While it is well known that the LNA {yields an exact prediction} for unimolecular reactions \cite{warren2006}, we derive closed-form expressions here for the leading order corrections to these power spectra. The latter are important for the study of networks that include arbitrary bimolecular or complex-elementary reactions. 
Firstly, we consider reaction networks that are composed of a single species. Secondly, we utilize the eigenvalue decomposition of the Jacobian of the rate equations to derive explicit expressions for the power spectrum in general multi-species networks. We illustrate our results by applying them to a simple network involving bursty protein production and dimerization.
Then, we investigate noise-induced oscillations, a phenomenon also known as coherence resonance \cite{ushakov2005,mckane2007}, in the Brusselator reaction scheme. These oscillations are captured neither by the deterministic rate equations nor by the corresponding LNA. In particular, we show that the theory presented here allows us to predict the dynamics of noise-induced oscillations over large ranges of parameters and molecule numbers. 

\section{The generating functional of the {CME}}

We consider a general reaction network which is confined in some volume $\Omega$, and which involves the interaction of $N$ distinct chemical species via $R$ chemical reactions of the type
 \begin{equation}
\label{eqn:reaction}
 s_{1j} X_1 + \ldots + s_{Nj} X_{N} \xrightarrow{k_j} \ r_{1j} X_{1} + \ldots +
r_{Nj} X_{N},
 \end{equation}
where $j$ is the reaction index running from $1$ to $R$, $X_i$ denotes chemical
species $i$, $k_j$ is the reaction rate of the $j^{th}$ reaction, and $s_{ij}$
and $r_{ij}$ are the stoichiometric coefficients. We denote the probability of observing the molecular population numbers $\vec{n}=({n_1,\ldots,n_N})^T$ at time $t$, for an ensemble initially prepared in state $\vec{n}_0$ at {some} previous time $t_0$, by ${P}(\vec{n},t|\vec{n}_0,t_0)$. Under well-mixed conditions, the latter can be shown to satisfy 
\begin{align}
\label{eqn:CME}
\dot{P}(\vec{n},t|\vec{n}_0,t_0)=
\Omega \sum\limits_{j=1}^R \left(\prod_{i=1}^{N}E^{-R_{ij}}_i-1\right)\hat{f}_j\left(\frac{\vec{n}}{\Omega},\Omega\right)P(\vec{n},t|\vec{n}_0,t_0),
\end{align}
which is called the CME \cite{vanKampen,gillespie1992}.
Here, $\matrix{R}$ is the stoichiometric matrix with {entries} $R_{ij} = r_{ij}-s_{ij}$, $\hat{f}_j\left(\frac{\vec{n}}{\Omega},\Omega\right)$ is the microscopic rate function given by the probability per unit time per unit volume for the $j^{th}$ reaction to occur \cite{gillespie2007}, and $E_i^{-R_{ij}}$ is the step operator which is defined by its action on a general function of molecular populations as $E_i^{-R_{ij}}g(n_1,\dots,n_i,\dots,n_N)=g(n_1,\dots,n_i-R_{ij},\dots,n_N)$
\cite{vanKampen}. Note that we underline matrices throughout the manuscript. 

\subsection{The exact functional integral of the CME}

Next, we describe how the generating functional containing the full information of the noise statistics can be expressed as a weighted sum over all possible trajectories of the underlying stochastic process. This representation is often referred to as the path-integral formulation and represents an exact statistic of the process. 

We therefore cast the CME, Eq.~(\ref{eqn:CME}), into the form $\dot{P}(\vec{n},t|\vec{n}_0,t_0)=\Omega\mathcal{L}(\hat{\vec{P}},\hat{\vec{Q}})P(\vec{n},t|\vec{n}_0,t_0)$ in which $\mathcal{L}$ is a function of the operators
\begin{align}
  \label{eqn:definePQ}
  \hat{\vec{Q}}=\frac{\vec{n}}{\Omega},\ \ \hat{\vec{P}}=-\nabla_{\vec{n}}^T,
\end{align}
where $\nabla_{\vec{n}}^T=(\partial_{n_1},\ldots,\partial_{n_N})$.
By the definition of the step operator $E_i$, its action on an analytic function $g$ can be written as $\prod_{i=1}^{N}E^{-R_{ij}}_i g(\vec{n})={\rm e}^{-(\nabla_{\vec{n}}^T \matrix{R})_j}g(\vec{n})={\rm e}^{(\hat{\vec{P}}\matrix{R})_j}g(\hat{\vec{Q}})$. 
Using these definitions, we can express the transition matrix $\mathcal{L}$ as the function
\begin{align}
\label{eqn:Lfunction}
\mathcal{L}(\hat{\vec{P}},\hat{\vec{Q}})= \sum\limits_j \left({\rm e}^{(\hat{\vec{P}} \matrix{R})_j }-1\right)\hat{f}_j(\hat{\vec{Q}},\Omega).
\end{align}
{The above expression is ``normally ordered'' in the sense that it has all derivatives $\hat{\vec{P}}$ to the left and the vector of molecular concentrations $\hat{\vec{Q}}$ to the right; {also, it }explicitly depends on $\Omega$.} The generating functional $Z(\vec{j},\vec{j}^\star)$ for general multi-time moments can then be written as the functional integral proposed by Calisto and Tirapegui \cite{calisto1993},
\begin{align}
Z(\vec{j},\vec{j}^\star) 
=&\left\langle \exp\left(\int_{t_0}^{t'} dt {\vec{j}^\star(t)\hat{\vec{P}}(t)+\vec{j}^T(t)\hat{\vec{Q}}(t)}\right)\right\rangle \notag\\
=&
\int \mathcal{D}\vec{Q}\mathcal{D}\vec{P} 
\exp  \int_{t_0}^{t'} \dd t \left( -\Omega\vec{P}\dot{\vec{Q}}+\Omega\mathcal{L}(\vec{P},\vec{Q})+\vec{j}^T\vec{Q}+\vec{P}\vec{j}^\star)\right)\delta(\vec{Q}(t_0)-\vec{\phi}_0).
\label{eqn:CMEpathintegral}
\end{align}
An explicit definition of the measure $\mathcal{D}\vec{Q}\mathcal{D}\vec{P}$ is provided in Appendix \ref{app:fi}, Eq.~(\ref{app:GF}), together with a detailed derivation of the {above generating functional}. Note that the Dirac delta function fixes the deterministic initial condition $\vec{Q}(t_0)=\vec{\phi}_0$ here. The merit of the functional integral approach lies in replacing the operators $\hat{\vec{Q}}$ and $\hat{\vec{P}}$ with a real-valued vector $\vec{Q}$ and a complex-valued vector $\vec{P}$, respectively, which commute. {Eq.~(\ref{eqn:CMEpathintegral})} determines the moments of the CME for all times and can hence be used, at least in principle, to define arbitrary time-correlation functions for the underlying stochastic process. 

{
Even though the functional integral in Eq.~(\ref{eqn:CMEpathintegral}) is an exact representation of the CME, it does not {provide} analytical insight into {these} correlation functions. The reason is that the integration in (\ref{eqn:CMEpathintegral}) cannot be carried out analytically in general because Eq.~(\ref{eqn:Lfunction}) {does} not represent a quadratic form, {which} typically prevents one from obtaining exact expressions for the intrinsic noise power spectrum. In the following section, we review a formal diagrammatic definition of {$m$-point} correlation functions{; then, we} develop simple approximations {for these functions on the basis of} the system size expansion.
}

\subsection{Correlation functions}

We now introduce the exact connected generating functional $W$ as 
\begin{align}
 \label{eqn:W}
 W(\vec{j},\vec{j}^\star) = \ln Z(\vec{j},\vec{j}^\star).
\end{align}
Note that the above definition is convenient computationally, since the first and second cumulants of $\hat{\vec{Q}}$ are given by the mean concentrations and the centered second moments, {respectively}. The $m$-point correlation functions can be defined in both the time and the frequency domain, as follows:
\begin{align}
 G_m(t_1,t_2,\dots,t_m)&=\prod_{k=1}^m \delta_{\vec{j}(t_k)} W(\vec{j},\vec{j}^\star) \notag\\
 &=\int_{-\infty}^\infty \left(\prod_{k=1}^m \frac{\dd \omega_k}{2\pi}{\rm e}^{i\omega_k t_k}\right) 
 \left[ \prod_{k=1}^m  \delta_{\vec{j}(\omega_k)} W(\vec{j},\vec{j}^\star) \right]_{(\vec{j},\vec{j}^\star)=(0,0)},
\end{align}
where $\delta_{\vec{j}(t)}\equiv\delta/\delta\vec{j}(t)$ is the functional derivative which, augmented by $\delta_{\vec{j}(t')}\vec{j}(t)=\delta(t-t')\matrix{1}$, follows the usual rules of differentiation. {Analogous definitions apply }in the frequency domain representation.
With the diagrammatic method in mind we can represent such $m$-point functions graphically. For example, the exact one-point function {corresponding to} the mean concentration may be represented symbolically by a diagram with one external line
\begin{align}
\begin{fmffile}{diagrams/onepoint}
\begin{fmfgraph*}(40,40)
  \fmfleft{P1} \fmfright{g1}
  \fmflabel{$t$}{P1}
  \fmf{plain,lab=$\omega$,l.side=left}{P1,g1}
\fmfblob{.3w}{g1}
\end{fmfgraph*}
\end{fmffile}
\label{eqn:onepoint}
\end{align}
where the blob, which remains to be specified, arises from differentiation of the cumulant generating functional $W$. 
Similarly, the exact two-point function can be represented by
\begin{align}
\label{eqn:twopointDiag}
\begin{fmffile}{diagrams/twopoint}
\begin{fmfgraph*}(80,50)
  \fmfleft{P1}
  \fmfright{P2}
  \fmflabel{$t_1$}{P1}
  \fmflabel{$t_2$}{P2}
  \fmf{plain,lab=$\omega_1$,l.side=left}{P1,g1}
  \fmf{plain,lab=$\omega_2$}{P2,g1}
\fmfblob{.3w}{g1}
\end{fmfgraph*}
\end{fmffile}
\end{align}
Under stationarity, the two-point correlation function only depends on the distance {between} the two time points. Hence,
\begin{align}
 G_2(t_1-t_2) 
&= 
 \left[ \delta_{\vec{j}(t_1)}\delta_{\vec{j}(t_2)} W(\vec{j},\vec{j}^\star) \right]_{(\vec{j},\vec{j}^\star)=(0,0)}\notag\\
&=
\int_{-\infty}^\infty \frac{\dd \omega}{2\pi}{\rm e}^{i\omega(t_1-t_2)} \delta_{\vec{j}(\omega)}\delta_{\vec{j}(-\omega)} W(\vec{j},\vec{j}^\star),
\end{align}
which implies that the frequency over the whole diagram in Eq.~(\ref{eqn:twopointDiag}) is conserved in the sense that $\omega=\omega_1=-\omega_2$. 
The intrinsic noise power spectrum is then defined by the Fourier transform of $G_2$,
\begin{align}
 \matrix{S}(\omega)=
 \left[
 \delta_{\vec{j}(\omega)}\delta_{\vec{j}(-\omega)} W(\vec{j},\vec{j}^\star) 
 \right]_{(\vec{j},\vec{j}^\star)=(0,0)}.
\end{align}
General stationary $m$-point functions can be defined analogously in the Fourier domain, via
\begin{align}
 G_m(\omega_1,\omega_2,\dots,\omega_{m-1}) = \delta\left(\sum_{k=1}^m \omega_k\right) \left[ \prod_{k=1}^m  \delta_{\vec{j}(\omega_k)} W(\vec{j},\vec{j}^\star) \right]_{(\vec{j},\vec{j}^\star)=(0,0)};
\end{align}
{hence,} the one-point function in Eq.~(\ref{eqn:onepoint}) contributes only at zero frequency, as expected. It is important to note that the derivatives of the cumulant generating functional $W$ involve only diagrams which are fully connected \cite{zinn2002}. Specifically, in the case of $m=3$, the stationary three-point function is represented by
 
\begin{align}
 \begin{fmffile}{diagrams/threepoint}
 \begin{fmfgraph*}(60,60)
   \fmfleft{P1}
   \fmfright{P3,P2}
   \fmflabel{$t_1$}{P1}
   \fmflabel{$t_2$}{P2}
   \fmflabel{$t_3$}{P3}
   \fmf{plain,lab=$\omega_1$,l.side=left}{P1,g1}
   \fmf{plain,lab=$\omega_2$,l.side=right}{P2,g1}
   \fmf{plain,lab=$\omega_1+\omega_2$,l.side=right}{P3,g1}
 \fmfblob{.3w}{g1}
 \end{fmfgraph*}
 \end{fmffile}\notag\\
\end{align}
which concludes the formal diagrammatic representation of these $m$-point correlations.

\section{System size expansion using Feynman rules}

In the limit as $\Omega\to\infty$, the variational principle tells us that the integral in Eq.~(\ref{eqn:CMEpathintegral}) must be dominated by the minimum of the exponential therein, which is determined from the variational equations
\begin{align}
\dot Q_\alpha&=\frac{\partial \mathcal{L}}{\partial P_\alpha}= \sum\limits_j {\rm e}^{(\vec{P} \matrix{R})_j}R_{\alpha j}f_j(\vec{Q}), \notag\\
\dot P_\alpha&=-\frac{\partial \mathcal{L}}{\partial Q_\alpha}= \sum\limits_j (1-{\rm e}^{(\vec{P} \matrix{R})_j})\frac{\partial}{\partial Q_\alpha}f_j(\vec{Q}).
\end{align}
Here, $f_j(\vec{Q})=\lim_{\Omega\to\infty}\hat{f}_j(\vec{Q},\Omega)$ denotes the macroscopic rate function. {The above equations {constitute} the exact description of well-mixed reaction networks described by Eq.~(\ref{eqn:reaction}) in the limit of infinite system size.} A particular solution of these is given by the macroscopic rate equations 
\begin{align}
 \label{eqn:REs}
 \frac{d \vec{\phi}}{dt}{} = \matrix{R}\vec{f}(\vec{\phi}),
\end{align}
which is found by letting $\vec{P}=0$ and $\vec{Q}=\vec{\phi}$; the latter represent the saddle point of the functional integral. We now proceed by expanding the exponent in Eq.~(\ref{eqn:CMEpathintegral}) around that point. {The system size expansion {is based on} the scaling assumption that the fluctuations about the macroscopic state $\vec\phi$ decay as the inverse square root of the system size
\begin{align}
\label{eqn:VKansatz}
\vec{Q}=\vec\phi+ \Omega^{-\frac12}\vec{q}, \ \
\vec{P}=\Omega^{-\frac12}\vec{p},
\end{align}
which is akin to van Kampen's ansatz \cite{vanKampen1976,vanKampen}. Note that applying {the latter} to Eq.~(\ref{eqn:CMEpathintegral}) provides a local approximation for the generating functional which is valid only in the vicinity of a stable fixed point of Eq.~(\ref{eqn:REs}). Here, we restrict our analysis to {that} case. ({The above scaling} may fail for instance near extinction thresholds or when the macroscopic equations exhibit multistability \cite{hanggi1984,moloney2006,kamenev2011}.) We also rescale} $\vec{j}=\sqrt{\Omega}\vec{J}$ and $\vec{j}^\star=\sqrt{\Omega}\vec{J}^\star$ correspondingly. The generating functional for $\vec{q}$ and $\vec{p}$ is now {given by}
\begin{align}
\label{eqn:preExpansion}
Z(\vec{J},\vec{J}^\star)&=\int \mathcal{D}\vec{q}\mathcal{D}\vec{p} 
\exp \int_{t_0}^{t'} \dd t \left( -\vec{p}\vec{\dot{q}}+(\Omega\mathcal{L}(\Omega^{-1/2}\vec{p},\vec{\phi}+\Omega^{-1/2}\vec{q})-\sqrt\Omega\vec{p}\vec{\dot\phi})+\vec{J}^T\vec{q}+\vec{p}\vec{J}^\star\right)\delta(\vec{q}(t_0)),
\end{align}
and is related to $Z(\vec{j},\vec{j}^\star)$ through
\begin{align}
Z(\vec{j},\vec{j}^\star)=\left(\exp\int_{t_0}^{t'} \dd t  \vec{\phi}^T \vec{j}\right) Z(\vec{J},\vec{J}^\star). 
\end{align}
It remains to expand $\mathcal{L}$ in powers of the system size $\Omega$. In order to make this dependence explicit, we expand the microscopic rate functions as
\begin{align}
 \label{eqn:microscopicRates}
 \vec{\hat{f}}(\vec{\phi},\Omega)= \sum_{n=0}^\infty\Omega^{-n}\vec{{f}}^{(n)}(\vec{\phi}),
\end{align}
{
where the first term is the macroscopic rate function $\vec{f}^{(0)}(\vec{\phi})\equiv\vec{{f}}(\vec{\phi})$ of the rate equations, Eq.~(\ref{eqn:REs}). It is important to note that the form of the above expansion is (i) necessary for the CME to have a \emph{macroscopic} (deterministic) limit \cite{vanKampen}, and (ii) {that it} is implied by the law of mass action derived from \emph{microscopic} considerations \cite{gillespie1992}. The expansion is now obtained along the same lines as in \cite{plosOne}, where the system size coefficients have been defined as
\begin{align}
\label{eqn:SSEcoeffs}
{D^{(n)}}_{ij..r}&=\sum\limits_\alpha R_{i\alpha}R_{j\alpha}\ldots R_{r\alpha}{f}^{(n)}_\alpha(\vec\phi),\notag\\
{J^{(n)}}_{ij..r}^{st..z}&=\frac{\partial}{\partial{\phi_s}}\frac{\partial}{\partial{\phi_t}}\dots\frac{\partial}{\partial{\phi_z}} {D^{(n)}}_{ij..r}.
\end{align}
Here, ${D^{(n)}}_{ij..r}$ {is the Taylor coefficient of order $\Omega^{-n}$} of the jump moments, evaluated at the macroscopic concentrations $\vec\phi$, ${J^{(n)}}_{ij..r}^{st..z}$ denotes their derivatives, and $R_{ij}$ are the elements of the stoichiometric matrix defined after Eq.~(\ref{eqn:CME}). In the following we omit the index $(n)$ in the case where $n=0$. In particular, $D_\alpha$ equals the right hand side of the macroscopic equations in (\ref{eqn:REs}), $J_\alpha^{\beta}$ denotes their Jacobian matrix, and $D_{\alpha\beta}$ is commonly referred to as the diffusion matrix.}
Expanding $\Omega\mathcal{L} $ in terms of $\Omega^{-1/2}$, one obtains
\begin{align}
\label{eqn:Lexpansion}
\Omega\mathcal{L} &=
\sum_{n=0}^\infty \sum_{j=0}^\infty \Omega^{-\frac{n-1}2-j}\mathcal{L}^{(j)}_n =\Omega^{\frac12}\mathcal{L}_0^{(0)} + \Omega^0\mathcal{L}_1^{(0)} + \mathcal{L}'.
\end{align}
{
The first term in the above expansion is $\mathcal{L}_0^{(0)}=p_\alpha D_\alpha$, which equals $\vec{p}\vec{\dot{\phi}}$. The next term of order $\Omega^0$ is given by the quadratic form
\begin{align}
 \label{eqn:FPoperator}
    \mathcal{L}_1^{(0)}&= {J}_\alpha^\beta p_\alpha q_\beta 
    + \frac{1}{2!} {D}_{\alpha\beta} p_{\alpha}p_\beta,
\end{align}
where $\mathcal{L}'$ collects all terms that contain inverse powers of $\Omega$. Here, we have adopted Einstein's summation convention in which repeated Greek indices are summed over. The first few terms of $\mathcal{L}'$ are given by}
\begin{align}
\label{eqn:Lprime}
\mathcal{L}' & =\Omega^{-\frac12} \mathcal{L}_0^{(1)} + \Omega^{-\frac12} \mathcal{L}_2^{(0)} + \Omega^{-1} \mathcal{L}_1^{(1)} + \Omega^{-1} \mathcal{L}_3^{(0)} +O(\Omega^{-\frac32}),
\end{align}
with
\begin{align}
\label{eqn:vertices}
\mathcal{L}_0^{(1)}&=
 p_\alpha {D}_{\alpha}^{(1)},
\notag\\
\mathcal{L}_1^{(1)}&=
 p_\alpha J^{(1)\beta}_{\alpha}  q_\beta 
+ \frac{1}{2!} p_\alpha p_\beta {D}^{(1)}_{\alpha\beta},
\notag\\
\mathcal{L}_2^{(0)}&=
\frac{1}{2!} p_\alpha{J}_{\alpha}^{\beta\gamma}  q_\beta q_\gamma 
+ \frac{1}{2!} p_\alpha p_\beta{J}_{\alpha\beta}^\gamma  q_\gamma 
+  \frac{1}{3!}p_\alpha p_\beta p_\gamma {D}_{\alpha \beta \gamma},  
\notag\\
\mathcal{L}_3^{(0)}&= \frac{1}{3!} p_\alpha  {J}_{\alpha\beta}^{\beta\gamma\delta}q_\beta q_\gamma q_\delta+
 \frac{1}{2!}\frac{1}{2!} p_\alpha p_\beta  {J}_{\alpha\beta}^{\gamma\delta}q_\gamma q_\delta
+  \frac{1}{3!}p_\alpha p_\beta p_\gamma{J}_{\alpha\beta\gamma}^{\delta} q_\delta
 + \frac{1}{4!}  p_\alpha p_\beta p_\gamma p_\delta{D}_{\alpha \beta \gamma \delta}.
\end{align}
{Note that by Eqs.~(\ref{eqn:FPoperator}) through (\ref{eqn:vertices}), the expansion of $\mathcal{L}$ in Eq.~(\ref{eqn:Lexpansion}) vanishes for $\vec{p}=0$ to each order in the system size, {thus ensuring conservation of probability; see also Appendix \ref{app:fi}.}} It then follows that in the limit as $\Omega\to\infty$, we can approximate the functional integral {$Z(\vec{J},\vec{J}^\star)$ defined in Eq.~(\ref{eqn:preExpansion})} by the Gaussian integral
\begin{align}
\label{eqn:Z0}
Z_0(\vec{J},\vec{J}^\star)
&=\lim_{\Omega\to\infty}Z(\vec{J},\vec{J}^\star)\notag\\
&=\int \mathcal{D}\vec{q}\mathcal{D}\vec{p}
\exp \int_{t_0}^{t'} \! \dd t \left( -\vec{p}\vec{\dot{q}}+\vec{p}\matrix{J}\vec{q}+\frac{1}{2}\vec{p}\matrix{D}\vec{p}^T+\vec{J}^T\vec{q}+\vec{p}\vec{J}^\star\right)\delta(\vec{q}(t_0)),
\end{align}
which is the LNA and {which} has also been obtained by Calisto and Tirapegui \cite{calisto1993}. Here, the matrix $[\matrix{J}]_{\alpha\beta}=J_\alpha^\beta$ is the Jacobian of the macroscopic rate equations, Eq.~(\ref{eqn:REs}), while $\matrix{D}=\matrix{R}\text{diag}(\vec{f})\matrix{R}^T$ denotes the diffusion matrix. It also {follows} that Eqs.~(\ref{eqn:Lprime}) and (\ref{eqn:vertices}) determine the leading order correction to the LNA.

\subsubsection*{Frequency-domain representation of the {LNA}}

Gaussian integrals can generally be evaluated exactly. We perform the integration explicitly in the frequency domain using the Fourier transform. Assuming stationary conditions, the integral will be independent of the initial and the final times, and hence, we can safely take $t_0 \to -\infty$ and $t'\to\infty$ in Eq.~(\ref{eqn:Z0}). The corresponding Fourier transform is then found from the frequency domain representations
\begin{align}
\label{eqn:ftransform}
\vec{q}(t)=\int\limits_{-\infty}^{\infty} \! \frac{\dd\omega}{2\pi} \,  {\rm e}^{i \omega t}\vec{q}(\omega),
\ \
\vec{p}(t)=\int\limits_{-\infty}^{\infty} \! \frac{\dd\omega}{2\pi} \, {\rm e}^{i \omega t}\vec{p}(\omega).
\end{align} 
Inserting the above expressions into Eq.~(\ref{eqn:Z0}), we have
\begin{align}
\label{eqn:Z_LNA}
&Z_0(\vec{J},\vec{J}^\star)= \int \mathcal{D}\vec{q}\mathcal{D}\vec{p}\, {\rm e}^{\mathcal{S}_0(\vec{p},\vec{q})} \delta(\vec{q}(t_0)),
\end{align}
where the exponential $\mathcal{S}_0$ is given by
\begin{align}
\label{eqn:S0}
\mathcal{S}_0 = \int\limits_{-\infty}^{\infty} \!  \frac{\dd\omega}{2\pi} &\left( -i\omega\vec{p}(\omega){\vec{q}(-\omega)}+
\vec{p}(\omega)\matrix{J}\vec{q}(-\omega)+\frac{1}{2}\vec{p}(\omega)\matrix{D}\vec{p}^T(-\omega) \right. 
\notag\\
 &\left. + {\rm e}^{-i\omega 0^+}\vec{J}^T(\omega)\vec{q}(-\omega)+\vec{p}(\omega)\vec{J}^\star(-\omega)\right).
\end{align}
Note that we have included a factor of ${\rm e}^{-i\omega 0^+}$ here, which accounts for {the Ito discretization scheme} of the time integral\footnote{The Ito discretization, which has been derived in Appendix \ref{app:fi}, Eq.~(\ref{app:GF}), has a contribution of $\sum_k \delta t \vec{J}^T_k\vec{q}_{k-1}$ in the exponential. Inserting Eq.~(\ref{eqn:ftransform}) and taking the the limit as $\delta t \to 0$ yields $\int {\dd\omega}/({2\pi}){\rm e}^{-i\omega 0^+}\vec{J}^T(\omega)\vec{q}(-\omega)$.}.
{
The Gaussian integral in Eq.~(\ref{eqn:Z_LNA}) may be performed directly by completing the square in the exponential. Alternatively, by partial integration, we can utilize the two relations}
\begin{align}
\int \mathcal{D}\vec{q}\mathcal{D}\vec{p} \, \frac{\delta}{\delta\vec{q}} {\rm e}^{\mathcal{S}_0(\vec{p},\vec{q})}  =0,
\int \mathcal{D}\vec{q}\mathcal{D}\vec{p} \, \frac{\delta}{\delta\vec{p}} {\rm e}^{\mathcal{S}_0(\vec{p},\vec{q})}  =0.
\end{align}
The first relation can be used to derive
\begin{align}
\label{eqn:1stcondition}
\frac{\delta Z_0}{\delta \vec{J}^\star(-\omega)}= \vec{J}^T(\omega) \matrix{F}(\omega) Z_0,
\end{align}
which, together with the second relation, yields
\begin{align}
\label{eqn:2ndcondition}
\frac{\delta Z_0}{\delta \vec{J}(\omega)}= (\matrix{F}(\omega) \vec{J}^\star(-\omega)+ \matrix{\Delta}(\omega) \vec{J}(-\omega)) Z_0,
\end{align}
where have made use of the following two definitions
\begin{align}
\label{eqn:fdt0}
\matrix{F}(\omega) &= (i\omega-\matrix{J})^{-1}{\rm e}^{-i\omega 0^+},
\\
\label{eqn:fdt}
\matrix{\Delta}(\omega) &= \matrix{F}(\omega)\matrix{D}\matrix{F}^\dagger(\omega).
\end{align}
The above pair of equations can be integrated to obtain the generating functional of the stationary statistics,
\begin{align}
\label{eqn:LNAfunctional}
Z_0(\vec{J},\vec{J}^\star)= \exp\int\limits_{-\infty}^{\infty} \!  \frac{\dd\omega}{2\pi} \left( \vec{J}(\omega)^T\matrix{F}(\omega)\vec{J}^\star(-\omega)+\frac{1}{2}\vec{J}^T(\omega)\matrix{\Delta}(\omega)\vec{J}(-\omega)
\right).
\end{align}
Note that we have omitted the explicit dependence on the initial condition here. Note also that $Z_0(0,0)=1$ due to conservation of probability. {It may be noted that, for the purposes of calculation, Eqs.~(\ref{eqn:fdt0}) and (\ref{eqn:fdt}) might be written in the eigenbasis of the Jacobian $\matrix{J}$. It is then clear that $\matrix{F}(\omega)$ and $\matrix{\Delta}(\omega)$ are well defined only when all {eigenvalues of the Jacobian} have negative real part, which is the case for a stable fixed point of the macroscopic rate equations, {as} required by van Kampen's ansatz; see also the discussion after Eq.~(\ref{eqn:VKansatz}).}

\subsubsection*{{LNA} of the power spectra and correlation functions}

Using the cumulant generating functional $W_0=\ln Z_0$, we now verify that the functional in Eq.~(\ref{eqn:LNAfunctional}) agrees with well-known results obtained from the LNA. The first cumulant is calculated as
\begin{align}
\langle \hat{\vec{q}}(\omega)\rangle_0=\delta_{\vec{J}(\omega)}W_0|_{(\vec{J},\vec{J}^\star)=(0,0)}=0.
\end{align}
Hence, it follows from Eq.~(\ref{eqn:VKansatz}) in combination with the LNA that the average concentrations are well predicted by the macroscopic rate equations. The second cumulant is given by the Fourier transform of the correlation function, as discussed earlier, and evaluates to
 \begin{align}
 &\langle \hat{\vec{q}}(\omega)\hat{\vec{q}}^T(-\omega)\rangle_{0}=\delta_{\vec{J}(\omega)}\delta_{\vec{J}(-\omega)}W_0|_{(\vec{J},\vec{J}^\star)=(0,0)}
=\matrix{\Delta}(\omega),
\end{align}
where $\matrix{\Delta}(\omega)$ is given by Eq.~(\ref{eqn:fdt}). The above is the well-known result for the spectral matrix, which can be found in \cite{gardiner}. 

We further have
$\delta_{\vec{J}(\omega)}\delta_{\vec{J}^\star(-\omega)}W_0|_{(\vec{J},\vec{J}^\star)=(0,0)}
=\matrix{F}(\omega)$, which is given by the Green's function of the linearized rate equations, as is seen by inverting the Fourier transform of Eq.~(\ref{eqn:fdt0}):
\begin{align}
\label{eqn:timedomain}
\matrix{F}(\tau)= \int \frac{\dd \omega}{2\pi}{\rm e}^{i\omega \tau} \matrix{F}(\omega)= H(\tau-0^+){\rm e}^{\matrix{J} \tau},
\end{align}
where $H(\tau)$ denotes the Heaviside step-function and we have made use of the fact that $(i\omega-\matrix{J})^{-1}=\int_0^\infty \dd s\,{\rm e}^{(\matrix{J}-i\omega)s}$ for asymptotically stable $\matrix{J}$. {Note that the factor of ${\rm e}^{-i\omega 0^+}$ in Eq.~(\ref{eqn:fdt0}) ensures that the contour of the Fourier integral has to be closed in the lower half of the complex plane {on} which $\matrix{F}(\tau=0)$ vanishes, {thus precluding} an instantaneous action of the linear response and, hence, implementing causality.} Similarly, the expression for $\matrix{\Delta}$ in the time domain, $\matrix{\Delta}(\tau)=\int \frac{\dd\omega}{2\pi}{\rm e}^{i \omega \tau} \matrix{\Delta}(\omega)$, can been obtained from
\begin{align}
\label{eqn:timedomain2}
\matrix{\Delta}(\tau)&=\matrix{F}(\tau)\matrix\sigma+\matrix\sigma\matrix{F}^T(-\tau),
\end{align}
where $\matrix{\Delta}(0)=\matrix\sigma$ is just the covariance matrix of $\vec{q}$ \cite{gardiner}.
{
This relation between the size of fluctuations $\matrix{\Delta}(\tau)$ and the linear response function $\matrix{F}(\tau)$, or equivalently its frequency domain representation, Eq.~(\ref{eqn:fdt}), is commonly obtained in dynamic perturbation theories of equilibrium and non-equilibrium steady states and referred to as the \emph{generalized fluctuation-dissipation theorem} \cite{keizer1987,tauber2014}.
}

\subsection{Setting up the perturbation expansion}

Next, we make use of a well-known trick \cite{srednicki} to express the exact generating functional in terms of the Gaussian integral $Z_0$, Eq.~(\ref{eqn:Z0}). For any analytic function $f(\hat{\vec{p}},\hat{\vec{q}})$, the Gaussian expectation value can be written as
\begin{align}
\label{eqn:GaussianExpectation}
\langle f(\hat{\vec{p}}(t),\hat{\vec{q}}(t'))\rangle_0 &=
\int \mathcal{D}\vec{q}\mathcal{D}\vec{p} \,
f(\vec{p}(t),\vec{q}(t'))
{\rm e}^{\mathcal{S}_0(\vec{p},\vec{q})}=f(\delta_{\vec{J}(t)},\delta_{\vec{J}^\star(t')})Z_0,
\end{align}
{where we have used Eq.~(\ref{eqn:Z_LNA}).} We thus define the function
\begin{align}
\label{eqn:K}
\mathcal{K}(\vec{p},\vec{q})&=\exp\int_{-\infty}^\infty \dd t \, \mathcal{L'}(\vec{p},\vec{q}),
\end{align}
and note that, using Eq.~(\ref{eqn:GaussianExpectation}), the exact generating functional can be expressed as
\begin{align}
\label{eqn:pertZ}
Z(\vec{J},\vec{J}^\star)&=\mathcal{K}(\delta_{\vec{J}^\star},\delta_{\vec{J}})Z_0(\vec{J},\vec{J}^\star)={\rm e}^{ \int
\dd t \, \mathcal{L'}(\delta_{\vec{J}^\star(t)},\delta_{\vec{J}(t)})}Z_0(\vec{J},\vec{J}^\star).
\end{align}
{Analogous expressions are common to dynamical perturbation theories \cite{srednicki,zinn2002}.}
The above expression represents an ideal starting point for the purposes of detailed calculation, since the exponential therein can be expanded in powers of the inverse square root of the system size. Using Eq.~(\ref{eqn:Lprime}) in Eq.~(\ref{eqn:K}), we find that, in the frequency domain, 
\begin{align}
\mathcal{K}(\delta_{\vec{J}^\star},\delta_{\vec{J}})&=
\exp \sum{}{'} \int_{-\infty}^\infty  \left(\prod_{k=1}^n \frac{\dd \omega_k}{2\pi}\right)
 \,
\delta\left(\sum_{k=1}^n \omega_k\right) \Omega^{-(\frac n2+j-1)}\mathcal{L}^{(j)}_n(\delta_{\vec{J}^\star},\delta_{\vec{J}}),
\end{align}
where the summation $\sum'$ is over all values $n>0$, $j\ge0$ satisfying $n+2j> 0$. This result can in principle be used to evaluate expectation values explicitly to any order in the system size expansion.

\subsection{Use of diagrams for explicit calculation}

\label{sec:UseOf}

In order to illustrate the {utility} of Feynman diagrams for the evaluation of expectation values, we exemplify here the calculation of the mean concentrations up to order $\Omega^{-1}$. The analysis is carried out in parallel both analytically and using the diagrammatic method. (Readers familiar with the technique can safely skip this section.)
We therefore expand
\begin{align}
\label{eqn:mean}
 \langle \hat{q}_i \rangle&=\left. \delta_{J_i(\omega)} Z\right|_{(\vec{J},\vec{J}^\star)=(0,0)}\notag\\
&=\Omega^{-\frac12}
\delta(\omega)\left.\delta_{J_i(\omega)}Z_0(\vec{J},\vec{J}^\star)\right|_{(\vec{J},\vec{J}^\star)=(0,0)}\notag\\
&+
\Omega^{-\frac12}\delta(\omega)\int_{-\infty}^\infty  \left(\prod_{k=1}^3 \frac{\dd \ell_k}{2\pi}\right)\,
\delta\left(\sum_{k=1}^3 \ell_k\right)
J_\alpha^{\beta\gamma}
\left(\delta_{J_\alpha^\star (\ell_1)}\delta_{J_\beta(\ell_2)}\delta_{J_\gamma(\ell_3)}\delta_{J_i(\omega)}
Z_0(\vec{J},\vec{J}^\star)
\right)_{(\vec{J},\vec{J}^\star)=(0,0)} \notag\\
&+
\Omega^{-\frac12}\delta(\omega)\int_{-\infty}^\infty  \left(\prod_{k=1}^3 \frac{\dd \ell_k}{2\pi}\right)\,
\delta\left(\sum_{k=1}^3 \ell_k\right)
J_{\alpha\beta}^{\gamma}
\left(\delta_{J_\alpha^\star(\ell_1)}\delta_{J_\beta^\star(\ell_2)}\delta_{J_\gamma(\ell_3)}\delta_{J_i(\omega)}
Z_0(\vec{J},\vec{J}^\star)
\right)_{(\vec{J},\vec{J}^\star)=(0,0)} \notag\\
&+
\Omega^{-\frac12}\delta(\omega)\int_{-\infty}^\infty  \left(\frac{\dd \ell}{2\pi}\right)\,
\delta\left(\ell\right)
D^{(1)}_\alpha 
\left( 
\delta_{J_\alpha^\star(\ell)}\delta_{J_i(\omega)} Z_0(\vec{J},\vec{J}^\star)
\right)_{(\vec{J},\vec{J}^\star)=(0,0)} \notag \\
&+
\Omega^{-\frac12}\delta(\omega)\int_{-\infty}^\infty  \left(\prod_{k=1}^3 \frac{\dd \ell_k}{2\pi}\right)\,
\delta\left(\sum_{k=1}^3 \ell_k\right)
D_{\alpha\beta\gamma}
\left(\delta_{J_\alpha^\star(\ell_1)}\delta_{J_\beta^\star(\ell_2)}\delta_{J_\gamma^\star(\ell_3)}\delta_{J_i(\omega)}
Z_0(\vec{J},\vec{J}^\star)
\right)_{(\vec{J},\vec{J}^\star)=(0,0)} \notag\\
&+ O(\Omega^{-1})
\end{align}
Since $\left.\delta_{J_i(\omega)}Z_0(\vec{J},\vec{J}^\star)\right|_{(\vec{J},\vec{J}^\star)=(0,0)}=0$, we are left with four integrals. These integrals can be evaluated analytically from the identities
{
\begin{align}
 \left(\delta_{\vec{J}(\omega)}\delta_{\vec{J}(\omega')}
Z_0(\vec{J},\vec{J}^\star)
\right)_{(\vec{J},\vec{J}^\star)=(0,0)}
&= \delta(\omega+\omega')\matrix{\Delta}(\omega), \notag\\
 \left(\delta_{\vec{J}^\star(\omega)}\delta_{\vec{J}(\omega')}
Z_0(\vec{J},\vec{J}^\star)
\right)_{(\vec{J},\vec{J}^\star)=(0,0)}
&= \delta(\omega+\omega') \matrix{F}(\omega),\notag\\
 \left(\delta_{\vec{J}^\star(\omega)}\delta_{\vec{J}^\star(\omega')}
Z_0(\vec{J},\vec{J}^\star)
\right)_{(\vec{J},\vec{J}^\star)=(0,0)}
&= 0,
\end{align}
in combination with Wick's theorem \cite{zinn2002,tauber2012}}, which states that the expectation value of a Gaussian {distribution} is given by the sum over all possible products of pairs:
\begin{align}
 \left(\delta_{{X}_1(\omega_1)}\ldots\delta_{{X}_{2n}(\omega_n)}Z_0\right)_{(\vec{J},\vec{J}^\star)=(0,0)}
= \sum_{\text{pairings}}
\left(\delta_{{X}_{i_1}} \delta_{{X}_{i_2}} Z_0\right)_{(\vec{J},\vec{J}^\star)=(0,0)}
\ldots
\left(\delta_{{X}_{i_{m-1}}} \delta_{{X}_{i_m}} Z_0\right)_{(\vec{J},\vec{J}^\star)=(0,0)}.
\end{align}
There is, however, a simpler way which is based on a representation of these integrals in diagrammatic form. We begin by assigning the line
$
\begin{fmffile}{diagrams/f_line}
\begin{fmfgraph}(50,5)
\fmfleft{i1}
\fmfright{o1}
\fmf{photon}{i1,o1}
\end{fmfgraph}
\end{fmffile}
$ to $\matrix{\Delta}(\omega)$
and the line
$
\begin{fmffile}{diagrams/d_line}
\begin{fmfgraph}(50,5)
\fmfleft{i1}
\fmfright{o1}
\fmf{fermion}{i1,o1}
\end{fmfgraph}
\end{fmffile}
$ to $\matrix{F}(\omega)$,
while denoting the matrices $J_{\alpha}^{\beta\gamma}$, $J_{\alpha\beta}^{\gamma}$, $D_{\alpha\beta\gamma}$ by
\,
$
\begin{fmffile}{diagrams/dot}
\begin{fmfgraph}(5,5)
\fmfleft{i1}
\fmfdot{i1}
\end{fmfgraph}
\end{fmffile}
$.
The second term in Eq.~(\ref{eqn:mean}) thus becomes
\begin{align}
\label{eqn:integral1}
\delta(\omega)\int_{-\infty}^\infty & \left(\prod_{k=1}^3 \frac{\dd \ell_k}{2\pi}\right)\,
\delta\left(\sum_{k=1}^3 \ell_k\right)
J_\alpha^{\beta\gamma}
\left(\delta_{J_\alpha^\star(\ell_1)}\delta_{J_\beta(\ell_2)}\delta_{J_\gamma(\ell_3)}\delta_{J_i(\omega)}
Z_0(\vec{J},\vec{J}^\star)
\right) 
\notag\\
=& \delta(\omega) [\matrix{F}(\omega)]_{i\alpha} \left(\frac{1}{2!} {J}_\alpha^{\beta\gamma}\int \frac{\dd \ell}{2 \pi} [\matrix{\Delta}(\ell)]_{\beta\gamma}\right)
+\delta(\omega) [\matrix{\Delta}(\omega)]_{i\alpha} \left(\frac{2}{2!} {J}_\alpha^{\beta\gamma}\int \frac{\dd \ell}{2 \pi} [\matrix{F}(\ell)]_{\beta\gamma}\right).
\notag\\
=&
\frac{1}{2!}
\begin{tabular}[c]{c}
\begin{fmffile}{diagrams/tadpole_0}
\begin{fmfgraph*}(40,40)
\fmfleft{i1} \fmfright{v1}
\fmf{fermion,label=$\omega$,tension=5}{i1,v1}
\fmf{photon,label=$\ell$,tension=0.7}{v1,v1}
\fmfdot{v1}
\fmflabel{${J}_\alpha^{\beta\gamma}$}{v1}
\fmflabel{$i$}{i1}
\end{fmfgraph*}
\end{fmffile}
\end{tabular}
\ \ \ \ \ \ \ \ \ \ \ \
+\frac{2}{2!}
\underbrace{
\begin{tabular}[c]{c}
\begin{fmffile}{diagrams/tadpole_1}
\begin{fmfgraph*}(40,40)
\fmfleft{i1} \fmfright{v1}
\fmf{photon,label=$\omega$,tension=5}{i1,v1}
\fmf{fermion,label=$\ell$,tension=0.7}{v1,v1}
\fmfdot{v1}
\fmflabel{${J}_\alpha^{\beta\gamma}$}{v1}
\fmflabel{$i$}{i1}
\end{fmfgraph*}
\end{fmffile}
\end{tabular}
}_{=0}
\end{align}
where we adopt the convention that $\ell$, being internal, has to be integrated out. Note that the second term appears twice due to the symmetry $J_\alpha^{\beta\gamma}=J_\alpha^{\gamma\beta}$, but that {the latter vanishes because $\int \frac{\dd \ell}{2 \pi} \matrix{F}(\ell)=0$, {by} Eq.~(\ref{eqn:timedomain}). The absence of closed loops {in} the linear response is thus a consequence of causality.
}

Similarly, we have
\begin{align}
\label{eqn:contrib1}
 \delta(\omega) & \int_{-\infty}^\infty \left(\prod_{k=1}^3 \frac{\dd \ell_k}{2\pi}\right)\,
\delta\left(\sum_{k=1}^3 \ell_k\right)
J_{\alpha\beta}^{\gamma}
\left(\delta_{J_\alpha^\star(\ell_1)}\delta_{J_\beta^\star(\ell_2)}\delta_{J_\gamma(\ell_3)}\delta_{J_i(\omega)}
Z_0(\vec{J},\vec{J}^\star)
\right) 
 \notag\\
&=\frac{2}{2!}
\begin{tabular}[c]{c}
\begin{fmffile}{diagrams/tadpole_2}
\begin{fmfgraph*}(40,40)
\fmfleft{i1} \fmfright{v1}
\fmf{fermion,label=$\omega$,tension=5}{i1,v1}
\fmf{fermion,label=$\ell$,tension=0.7}{v1,v1}
\fmfdot{v1}
\fmflabel{${J}_{\alpha\beta}^{\gamma}$}{v1}
\fmflabel{$i$}{i1}
\end{fmfgraph*}
\end{fmffile}
\end{tabular}
\ \ \ \ \ \ \ \ \ \ \ \
+\frac{1}{2!}
\begin{tabular}[c]{c}
\begin{fmffile}{diagrams/tadpole_3}
\begin{fmfgraph*}(40,40)
\fmfleft{i1} \fmfright{v1}
\fmf{photon,label=$\omega$,tension=5}{i1,v1}
\fmf{fermion,label=$\ell$,tension=0.7}{v1,v1}
\fmfdot{v1}
\fmflabel{${J}_{\alpha\beta}^{\gamma}$}{v1}
\fmflabel{$i$}{i1}
\end{fmfgraph*}
\end{fmffile}
\end{tabular} 
\notag\\ 
&=0 
\end{align}
and
\begin{align}
\label{eqn:contrib2}
\delta(\omega)&\int_{-\infty}^\infty \left( \frac{\dd \ell}{2\pi}\right)\,
\delta\left( \ell\right)
D^{(1)}_\alpha 
\left( 
\delta_{J_\alpha^\star(\ell)}\delta_{J_i(\omega)} Z_0(\vec{J},\vec{J}^\star)
\right) \notag\\
&= -\Omega^{-\frac12} \matrix{J}^{-1}_{i\alpha} D_\alpha^{(1)} 
= 
\begin{tabular}[c]{c}
\begin{fmffile}{diagrams/C0}
\begin{fmfgraph*}(40,40)
\fmfleft{i1} \fmfright{v1}
\fmf{plain,label=$\omega$,tension=5}{i1,v1}
\fmfct{v1}
\fmfv{lab=${D}_\alpha^{{(1)}}$,lab.d=0.25w,lab.a=0}{v1}
\fmflabel{$i$}{i1}
\end{fmfgraph*}
\end{fmffile}
\end{tabular}
  \end{align}
where we have introduced an extra vertex \,\,\,$
\begin{fmffile}{diagrams/cross}
\begin{fmfgraph}(2,2)
\fmfleft{i1}
\fmfct{i1}
\end{fmfgraph}
\end{fmffile}$ for ${D}_\alpha^{{(1)}}$ which attaches only one external line.
Combining Eq.~(\ref{eqn:mean}) with Eqs.~(\ref{eqn:integral1}) through (\ref{eqn:contrib2}), we find that, up to order $\Omega^{-1/2}$, $\langle \hat{q}_i\rangle$ is then simply the sum of the following two diagrams:
\begin{align}
\langle \hat{q}_i(\omega) \rangle=\Omega^{-\frac12}\left(\frac{1}{2!}
\begin{tabular}[c]{c}
\begin{fmffile}{diagrams/tadpole}
\begin{fmfgraph*}(40,40)
\fmfleft{i1} \fmfright{v1}
\fmf{plain,label=$\omega$,tension=5}{i1,v1}
\fmf{photon,label=$\ell$,tension=0.7}{v1,v1}
\fmfdot{v1}
\fmflabel{${J}_\alpha^{\beta\gamma}$}{v1}
\fmflabel{$i$}{i1}
\end{fmfgraph*}
\end{fmffile}
\end{tabular}
\qquad \qquad  +
\begin{tabular}[c]{c}
\begin{fmffile}{diagrams/C0}
\begin{fmfgraph*}(40,40)
\fmfleft{i1} \fmfright{v1}
\fmf{plain,label=$\omega$,tension=5}{i1,v1}
\fmfct{v1}
\fmfv{lab=${D}_\alpha^{{(1)}}$,lab.d=0.25w,lab.a=0}{v1}
\fmflabel{$i$}{i1}
\end{fmfgraph*}
\end{fmffile}
\end{tabular}
\qquad
\right)
+ O(\Omega^{-1}).
\label{diag:one_point}
\end{align}
These diagrams are known as tadpoles, and lead to the corrections to the mean concentrations predicted by the rate equations. {Note that the above sum of diagrams differs from a pure loop expansion by the second diagram, the reason being the explicit dependence of the microscopic rate functions on the system size $\Omega$. In the following, we denote these extra vertices, more specifically those with $n\neq 0$ in Eq.~(\ref{eqn:SSEcoeffs}), by \,
$\begin{fmffile}{diagrams/cross}
\begin{fmfgraph}(2,2)
\fmfleft{i1}
\fmfct{i1}
\end{fmfgraph}
\end{fmffile}$.} Noting, moreover, that the closed loop above corresponds to $\matrix{\sigma}=\int \frac{\dd \ell}{2 \pi} \matrix{\Delta}(\ell)$, we can write
\begin{align}
 \label{eqn:EMRE}
 \langle \hat{q}_i \rangle 
 =-\Omega^{-\frac12} \matrix{J}^{-1}_{i\alpha}\left( \frac{1}{2!} J_\alpha^{\beta\gamma} \sigma_{\beta\gamma} + D_\alpha^{(1)}\right)\equiv \Omega^{-\frac12} \epsilon_i.
\end{align}
We may now conclude that the average concentration is given by
\begin{align}
 \label{eqn:EMRE2}
 \left\langle \frac{n_i}{\Omega} \right\rangle 
 = \vec{\phi} -\frac{1}{\Omega} \matrix{J}^{-1}_{i\alpha}\left( \frac{1}{2!} J_\alpha^{\beta\gamma} \sigma_{\beta\gamma} + D_\alpha^{(1)}\right)+O(\Omega^{-2}),
\end{align}
to order $\Omega^{-1}$. 
The same result has been obtained from EMREs (Effective Mesoscopic Rate Equations) by Grima \cite{grima2010} for the case of purely elementary reactions, albeit on the basis of a truncation of the partial differential equation underlying van Kampen's system size expansion to order $\Omega^{-1/2}$. As we have seen above, the direct computation of these corrections can be quite involved. 
In the next section, we summarize a set of rules, commonly known as Feynman rules, which greatly simplify the required calculations to a merely combinatorial task, and which hence eliminate the need to invoke Wick's theorem directly.

\subsection{The Feynman rules for the correlation functions}

\label{sec:rules}

The above observations can be summarized by the following set of rules:

\begin{enumerate}[(i)]
 \item To calculate the $m$-point cumulant correlation function, draw $m$ lines (called external lines). 
 \item Pick a number of symbols either from $D_{i_1..i_n}$, $J_{i_1\dots}^{\dots i_n}$ or $D^{(s)}_{i_1..i_n}$, $J_{i_1\dots}^{(s)\dots i_n}$. {Denote the former by 
\,\,\,$\begin{fmffile}{diagrams/dot}
\begin{fmfgraph}(5,5)
\fmfleft{i1}
\fmfdot{i1}
\end{fmfgraph}
\end{fmffile}$, and the latter by
\,\,\,$
\begin{fmffile}{diagrams/cross}
\begin{fmfgraph}(5,5)
\fmfleft{i1}
\fmfct{i1}
\end{fmfgraph}
\end{fmffile}$.} A symbol with $n$ indices is represented by an $n$-vertex (a node in which exactly $n$ lines meet). 
 \item Leave one end of each external line free and attach the other end to a vertex index. Draw lines between the remaining vertex indices such that the diagram is connected.
 \item For each external line account with a factor of $\Omega^{-1/2}$, for each vertex $D_{i_1..i_n}$ or $J_{i_1\dots}^{\dots i_n}$ account {with} a factor of $\Omega^{-(n-2)/2}$, and for each vertex $D^{(s)}_{i_1..i_n}$ or $J_{i_1\dots}^{(s)\dots i_n}$ account {with} a factor of $\Omega^{-s-(n-2)/2}$. The order in the system size expansion to which the diagram contributes is the product of these factors. 
\item Repeat steps (i)-(iv) and draw all possible diagrams up to the desired order in the system size expansion.
  \item To each line that connects two upper vertex indices, assign a propagator 
$\begin{fmffile}{diagrams/f_line}
\begin{fmfgraph}(50,5)
\fmfleft{i1}
\fmfright{o1}
\fmf{photon}{i1,o1}
\end{fmfgraph}
\end{fmffile}$. 
To each line that connects an upper and a lower vertex index, assign a propagator $
\begin{fmffile}{diagrams/d_line}
\begin{fmfgraph}(50,5)
\fmfleft{i1}
\fmfright{o1}
\fmf{fermion}{i1,o1}
\end{fmfgraph}
\end{fmffile}$. Note that there are no lines connecting two lower vertex indices.

 \item Assign 
$\begin{fmffile}{diagrams/d_line}
\begin{fmfgraph}(50,5)
\fmfleft{i1}
\fmfright{o1}
\fmf{fermion}{i1,o1}
\end{fmfgraph}
\end{fmffile}$ to external lines attached to an lower vertex index, and otherwise $\begin{fmffile}{diagrams/f_line}
\begin{fmfgraph}(50,5)
\fmfleft{i1}
\fmfright{o1}
\fmf{photon}{i1,o1}
\end{fmfgraph}
\end{fmffile}$.
 \item Assign a frequency to each line, and conserve frequency at each vertex. Conserve also total frequency over the entire diagram.
 \item The value of each diagram is given by the product of the factors associated with all lines and vertices. To each line
$\begin{fmffile}{diagrams/f_line}
\begin{fmfgraph}(50,5)
\fmfleft{i1}
\fmfright{o1}
\fmf{photon}{i1,o1}
\end{fmfgraph}
\end{fmffile}$ with frequency $\omega$, assign the factor $\matrix{\Delta}(\omega)$. To each line $
\begin{fmffile}{diagrams/d_line}
\begin{fmfgraph}(50,5)
\fmfleft{i1}
\fmfright{o1}
\fmf{fermion}{i1,o1}
\end{fmfgraph}
\end{fmffile}$ with frequency $\omega$, assign the factor $\matrix{F}(\omega)$.
 \item A diagram with $L$ internal loops will have $L$ internal frequencies that are not fixed by the requirement for frequency conservation. Integrate over all frequencies, with measure $\dd \ell_i/(2\pi)$.
 \item Sum over all internal indices, accounting for the symmetry properties of the given diagram. The corresponding symmetry factor equals the number of ways in which the vertices in a diagram can be connected, divided by the symmetry of the vertex factors. {The latter is given by $n_u!n_d!$, since the vertex factor $J_{i_1\dots i_{n_d}}^{j_1\dots j_{n_u}}$ is invariant against} permutation of all upper or lower indices.
 \item In conclusion, the $m$-point correlation function equals the sum over all connected diagrams with $m$ external lines.

\end{enumerate}

\section{Corrections to the {LNA} of the power spectra}
\begin{figure}[t]
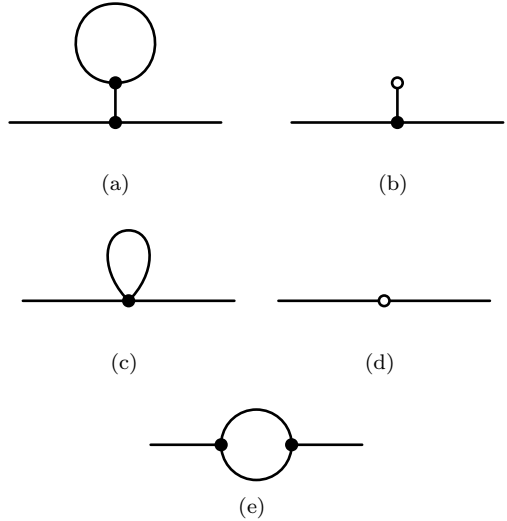

  \centering
  \include{topologies} 
  \caption{\textbf{Construction of two-point diagrams.} Distinct topologies of two-point diagrams contributing to the intrinsic noise power spectrum to order $\Omega^{-2}$. Since each external line contributes a factor of $\Omega^{-1/2}$ we only need to consider diagrams comprising a single vertex of order $\Omega^{-1}$ or combinations of two vertices {which each contribute} a factor of $\Omega^{-1/2}$. {Note that the system size expansion involves diagrams (b) and (d) that differ from a pure loop expansion.}}
  \label{fig:topologies}
\end{figure}

We now demonstrate the diagrammatic technique {by evaluating} the power spectrum to order $\Omega^{-2}$, i.e., including the next term beyond the LNA, which is important for reaction networks involving bimolecular reactions. The calculation of the requisite corrections is greatly facilitated by the set of Feynman rules presented in the previous section. 
The power spectrum is simply obtained as the sum of all diagrams with two external lines that include products of vertices of order $\Omega^{-2}$. It is clear {from Eqs.~(\ref{eqn:Lprime}) and (\ref{eqn:vertices})} that, to this order, we only need to consider diagrams with at most two vertices.

{Since the {system size expansion} involves non-zero corrections to the mean concentrations, as derived in Section \ref{sec:UseOf}, it is important to ensure that the expressions for the power spectrum of fluctuations are indeed centered around the corrected mean concentrations. Note that the generating functionals for the centered concentration fluctuations $(\hat{\vec{Q}}-\langle\hat{\vec{Q}}\rangle)$ and the deviations from the rate equations $\hat{\vec{q}}$, Eq.~(\ref{eqn:pertZ}), are related by the exponential factor ${\rm e}^{\Omega^{1/2}\int dt' \vec{J}(t') \langle\vec{q}(t')\rangle}$. Thus, it follows that the sequence of cumulants generated by their logarithms agrees beyond the first one, and, hence, {that} these generate an equivalent series of connected diagrams which is used in the following. 
}

In order to simplify the construction of these diagrams, we divide them into the three basic topologies shown in Figs.~\ref{fig:topologies}(a) and (b), which are denoted as tadpoles; Figs.~\ref{fig:topologies}(c) and (d), {which we denote as snails, but which are also referred to as tadpoles in the literature;} and Fig.~\ref{fig:topologies}(e), which are denoted as loop diagrams, respectively. We have already encountered tadpoles in Section~\ref{sec:UseOf}. Denoting the contribution from tadpole diagrams by $\mathscr{T}$, the contribution from snail diagrams by $\mathscr{S}$, and the contribution from loop diagrams by $\mathscr{L}$, we can express the spectral matrix $\matrix{S}$ as
\begin{align}
 \matrix{S}(\omega)
 = \frac{1}{\Omega}\Delta(\omega) 
 + \frac{1}{\Omega^2}\mathscr{T}(\omega) 
 + \frac{1}{\Omega^2}\mathscr{S}(\omega)
 +\frac{1}{\Omega^2}\mathscr{L}(\omega) + O(\Omega^{-3});
\end{align}
here, the last two terms determine the power spectrum to order $\Omega^{-2}$. We now {outline} the explicit evaluation, in turn, of these diagrams. 

\subsection{Tadpoles, and all that\dots}

\begin{figure}[t]
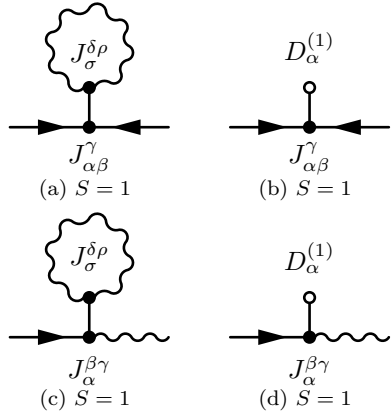

  \centering
  \include{tadpoles} 
  \caption{\textbf{Tadpole diagrams}. All tadpole diagrams contributing to order $\Omega^{-2}$ to the intrinsic noise power spectrum. Note that {none} of these have internal symmetries; hence, the symmetry factor is $S=1$ throughout.}
  \label{fig:tadpoles}
\end{figure}

In Fig.~\ref{fig:tadpoles}, we list all possible diagrams stemming from tadpoles. These contain one subdiagram which has already been evaluated in Section~\ref{sec:UseOf}. The sum of the diagrams in (a) and (b) is then given by
\\ \\ \\
\begin{align}
\label{eqn:T1}
\notag
\mathscr{T}_1(\omega)=
\frac{1}{2!}&
\begin{tabular}[c]{c}
\begin{fmffile}{diagrams/D5}
\begin{fmfgraph*}(60,30)
\fmfright{o1} \fmfleft{i1} \fmftop{v2}
\fmf{fermion,tension=5}{i1,v1}
\fmf{fermion,tension=5}{o1,v1}
\fmf{plain,tension=0}{v1,v2}
\fmf{photon,tension=1/3}{v2,v2}
\fmfdot{v1,v2}
\fmfv{lab=${J}_{\sigma}^{\delta\rho}$,lab.d=-0.3w,lab.a=-90}{v2}
\fmfv{lab=${J}_{\alpha\beta}^{\gamma}$,lab.d=-0.3w,lab.a=90}{v1}
\end{fmfgraph*}
\end{fmffile}
\end{tabular}
+
\frac{1}{2!}
\begin{tabular}[c]{c}
\begin{fmffile}{diagrams/TC1}
\begin{fmfgraph*}(60,30)
\fmfright{o1} \fmfleft{i1} \fmftop{v2}
\fmf{fermion,tension=5}{i1,v1}
\fmf{fermion,tension=5}{o1,v1}
\fmf{plain,tension=0}{v1,v2}
\fmfdot{v1}
\fmfv{lab=${D}_{\alpha}^{(1)}$,lab.d=-0.35w,lab.a=-90}{v2}
\fmfv{lab=${J}_{\alpha\beta}^{\gamma}$,lab.d=-0.3w,lab.a=90}{v1}
\end{fmfgraph*}
\end{fmffile}
\end{tabular}+\hc\\
&= \left(\frac{1}{ 2!}\right){F_{i\alpha } (\omega)} {J}_{\alpha\beta}^{\gamma} \epsilon_\gamma F_{\beta j}^\dagger (\omega)+\hc
\end{align}
Note that, by including also the Hermitian conjugate ($\hc$) of the above diagram, we have accounted for the contribution from an equivalent time-reversed diagram. Note also that $\epsilon_\gamma$ is defined by Eq.~(\ref{eqn:EMRE}).
Analogously, for the diagrams in Figs.~\ref{fig:tadpoles}(c) and (d), we find
\\ \qquad \\ \qquad \\
\begin{align}
\label{eqn:T2}
\mathscr{T}_2(\omega)=
\notag
\frac{1}{2!}&
\begin{tabular}[c]{c}
\begin{fmffile}{diagrams/D1}
\begin{fmfgraph*}(60,30)
\fmfright{o1} \fmfleft{i1} \fmftop{v2}
\fmf{fermion,tension=5}{i1,v1}
\fmf{photon,tension=5}{v1,o1}
\fmf{plain,tension=0}{v1,v2}
\fmf{photon,tension=1/3}{v2,v2}
\fmfdot{v1,v2}
\fmfv{lab=${J}_{\sigma}^{\delta\rho}$,lab.d=-0.35w,lab.a=-90}{v2}
\fmfv{lab=${J}_{\alpha}^{\beta\gamma}$,lab.d=-0.3w,lab.a=90}{v1}
\end{fmfgraph*}
\end{fmffile}
\end{tabular}
+
\frac{1}{2!}
\begin{tabular}[c]{c}
\begin{fmffile}{diagrams/TC2}
\begin{fmfgraph*}(60,30)
\fmfright{o1} \fmfleft{i1} \fmftop{v2}
\fmf{fermion,tension=5}{i1,v1}
\fmf{photon,tension=5}{v1,o1}
\fmf{plain,tension=0}{v1,v2}
\fmfct{v2}
\fmfdot{v1}
\fmfv{lab=$D_\alpha^{(1)}$,lab.d=-0.35w,lab.a=-90}{v2}
\fmfv{lab=$J_\alpha^{\beta\gamma}$,lab.d=-0.3w,lab.a=90}{v1}
\end{fmfgraph*}
\end{fmffile}
\end{tabular} +\hc\\
& = \left(\frac{1}{ 2!}\right){F_{i\alpha } (\omega)} {J}_\alpha^{\beta\gamma} \epsilon_\gamma \Delta_{\beta j} (\omega)+\hc
\end{align}
Finally, we denote the sum over the diagrams shown in Fig.~\ref{fig:tadpoles} by
\begin{align}
 \mathscr{T}(\omega)=\mathscr{T}_1(\omega) + \mathscr{T}_2(\omega).
\end{align}

\subsection{About snails\dots}

\begin{figure}
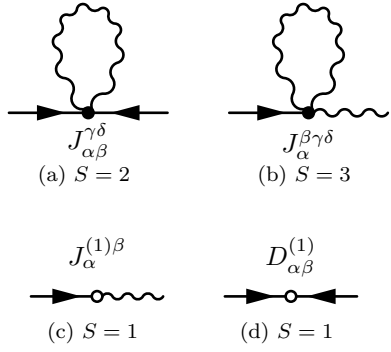

  \centering
  \include{snails}
  \caption{\textbf{Snail diagrams:}
  All snail diagrams contributing to order $\Omega^{-2}$ to the intrinsic noise power spectrum. The symmetry factors ($S$) follow by considering the ways in which the right external line can be attached to the corresponding vertex.
  } 
  \label{fig:snails}
\end{figure}

The next topology, Fig.~\ref{fig:topologies}(c), contains only a 4-single vertex where two upper indices are contracted in a loop. We have already found that such a loop equals the correlation matrix $\matrix{\sigma}$. The resulting diagrams are given in Figs.~\ref{fig:snails}(a) and (b),
implying
\begin{align}
 \label{eqn:T34}
 \mathscr{S}_1(\omega)&= \left( \frac{2}{2!2!}\right) {F_{i\alpha } (\omega)} J_{\alpha\beta}^{\gamma\delta} \sigma_{\gamma\delta} F_{\beta j}^\dagger(\omega)+\hc, \\
 \mathscr{S}_2(\omega)&= \left( \frac{3}{3!}\right) {F_{i\alpha } (\omega)} J_{\alpha}^{\beta\gamma\delta} \sigma_{\gamma\delta} \Delta_{\beta j}^\dagger(\omega)+\hc
\end{align}
Note that the diagram in Fig.~\ref{fig:snails}(b) is non-zero only when one considers reactions with non-elementary propensity functions, e.g., trimolecular reactions or those of Michaelis-Menten type \cite{thomas2011}.

We progress to the topology shown in Fig.~\ref{fig:topologies}(d), which is resolved by a simple matrix multiplication. The two possible diagrams are generated from 2-vertices, as shown in Figs.~\ref{fig:snails} (c) and (d). The result is 
\begin{align}
 \label{eqn:T56}
 \mathscr{S}_3(\omega)&=  {F_{i\alpha } (\omega)} J_{\alpha}^{(1)\beta}\Delta_{\beta j}^\dagger(\omega) +\hc,\\
 \mathscr{S}_4(\omega)&=  \frac{1}{2!} {F_{i\alpha } (\omega)} D_{\alpha\beta}^{(1)} F_{\beta j}^\dagger(\omega)+\hc, 
\end{align}
We denote the sum over all these diagrams by
\begin{align}
 \mathscr{S}(\omega)=\sum_{n=1}^4 \mathscr{S}_n(\omega),
\end{align}
which concludes the evaluation of the diagrams shown in Fig.~\ref{fig:snails}.

\subsection{Concerning loop diagrams}
\label{sec:loopdiagrams}

\begin{figure}
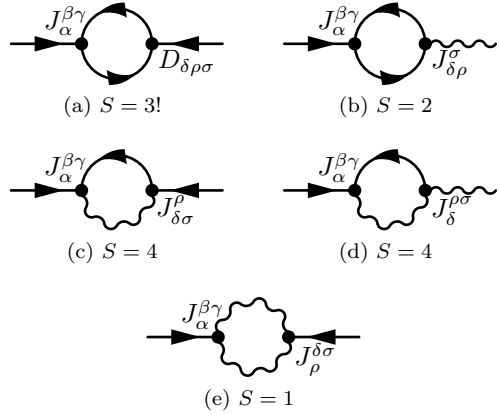

  \centering
  \include{loops}
  \caption{\textbf{Loop diagrams}. All diagrams that contribute to order $\Omega^{-2}$ to the intrinsic noise power spectrum, and that contain a frequency loop integral. The symmetry factors ($S$) follow by considering the ways in the right external line can be attached to the corresponding vertex, times the permutations of all internal lines. Note that in (e), we have only accounted for half of these permutations, since the diagram is self-adjoint.} 
  \label{fig:loops}
\end{figure}

The evaluation of all diagrams considered thus far has been possible by simple matrix multiplication. However, to evaluate the diagrams in the topology in Fig.~\ref{fig:topologies}(d), namely those which contain a loop formed by connecting two 3-vertices, one has to calculate explicitly the integrals occurring therein. The relevant diagrams are listed in Fig.~\ref{fig:loops}, all of which consist of a line $\matrix{F}$ connected to the 3-vertex $J_\alpha^{\beta\gamma}$, and are hence non-zero only for networks involving bimolecular reactions. 
We can then write the overall contribution from each diagram in the general form
\begin{align}
 \mathscr{L}_n(\omega) = \frac{1}{2}F_{i\alpha}(\omega) J_\alpha^{\beta\gamma}\left[\vec{\matrix{V}}^{j}_{(n)}(\omega)\right]_{\beta\gamma}+\hc,
\end{align}
where we have defined the vertex functions $\vec{\matrix{V}}^{j}_{(n)}(\omega)$, which are effectively 3-tensors. As we shall see, the latter consist of a 3-vertex with the frequency loop integrated out. 

\subsubsection*{Evaluation of vertex functions}

Let us study in detail the vertex function $\vec{\matrix{V}}^{j}_{(1)}(\omega)$ that is defined by the right hand side of the first diagram shown in Fig.~\ref{fig:loops}(e). The latter is simply the diagram with the vertex $J_\alpha^{\beta\gamma}$ removed, and can be written as
\begin{align}
\left[\vec{\matrix{V}}^{j}_{(1)}(\omega)\right]_{\beta\gamma} &\equiv \frac{3\times 2}{3!} \ \
\begin{tabular}[c]{c}
\begin{fmffile}{diagrams/D4vertex}
\begin{fmfgraph*}(40,40)
\fmfright{i1}
\fmfleft{o1,o2}
\fmf{fermion,lab=$\omega$,lab.dist=-0.3w}{i1,v}
\fmf{fermion,label=$\ell$}{o1,v}
\fmf{fermion,label=$\omega+\ell$}{v,o2}
\fmfdot{v}
\fmfv{lab=$j$}{i1}
\fmfv{lab=$\gamma$,lab.d=0.02w}{o1}
\fmfv{lab=$\beta$,lab.d=0.02w}{o2}
\fmfv{lab=$D_{\rho\delta\sigma}$,lab.dist=-0.75w,lab.a=0}{v}
\end{fmfgraph*}
\end{fmffile}
\end{tabular}
\notag\\
&=\int_{-\infty}^{\infty} \frac{\dd \ell}{2\pi}
F_{\beta\rho}(\omega+\ell)
\left( D_{\rho\delta\sigma} F_{\sigma j}^\dagger(\omega)\right)
 F_{\delta\gamma}^\dagger(\ell),
\end{align}
where the symmetry factor has been obtained via the considerations in the caption of Fig.~\ref{fig:loops}. We now evaluate the integral by inserting the Fourier transform of $\matrix{F}(\omega)$, Eq.~(\ref{eqn:timedomain}), into the above equation to show that its solution can be obtained by solving a linear matrix equation. We start by fixing the external index $j$ and write down the 3-vertex matrix with the frequency loop integrated out:
\begin{align}
 \label{eqn:vertexintegral}
 \vec{\matrix{V}}_{(1)}^j(\omega)
&=\int_{-\infty}^{\infty} \frac{\dd \ell}{2\pi}\matrix{F}(\omega+\ell)\matrix{M}_{(1)}^j(\omega)\matrix{F}^\dagger(\ell) \notag\\
&=\int_{0}^{\infty} \dd \tau {\rm e}^{(\matrix{J}-i\omega)\tau} \matrix{M}_{(1)}^j(\omega) {\rm e}^{\matrix{J}^T\tau},
\end{align}
where we have set $[\matrix{M}_{(1)}^j]_{\rho\delta}=D_{\rho\delta\sigma} F_{\sigma j}^\dagger(\omega)$.
As is shown in Appendix \ref{app:sylvester}, one can verify that, in this form, $\vec{\matrix{V}}_{(1)}^j$ satisfies Sylvester's matrix equation
\begin{align}
 \mathcal{S} \vec{\matrix{V}}_{(1)}^j=\matrix{M}_{(1)}^j (\omega),
 \label{eqn:loop_vertices0}
\end{align}
where $\mathcal{S}$ represents the linear operator
\begin{align}
 \label{eqn:Soperator}
 \mathcal{S}(\bullet)\equiv(\matrix{J}-i\omega)(\bullet)+(\bullet)\matrix{J}^T.
\end{align}
It is useful to define also the adjoint of $\mathcal{S}$ through
\begin{align}
 \label{eqn:Sadjoint}
 (\bullet)\mathcal{S}^\dagger\equiv(\matrix{J}+i\omega)(\bullet)+(\bullet)\matrix{J}^T.
\end{align}

\subsubsection*{Sum of loop diagrams}

The {vertex functions corresponding to} the diagrams shown in Figs.~\ref{fig:loops}(b) through (d) {can be evaluated in an analogous fashion, as demonstrated in Appendix \ref{app:vertexfunctions}, and are given by
\begin{align}
 \left[\mathcal{S} \vec{\matrix{V}}_{(1)}^j\right]_{\beta\gamma}&=D_{\beta\gamma\sigma} F_{\sigma j}^\dagger(\omega),
\notag
\\
 \left[\mathcal{S} \vec{\matrix{V}}_{(2)}^j\right]_{\beta\gamma}&=J_{\beta\gamma}^\delta \Delta_{\delta j}(\omega),
\notag
\\
 \left[\mathcal{S} \vec{\matrix{V}}_{(3)}^j\right]_{\beta\gamma}&=2\, J_{\beta\sigma}^\gamma F_{\sigma j}^\dagger(\omega),
\notag
\\
 \left[\mathcal{S} \vec{\matrix{V}}_{(4)}^j\right]_{\beta\gamma}&=2\, \sigma_{\beta\rho} J_\gamma^{\rho\delta} \Delta_{\delta j}(\omega),
 \label{eqn:loop_vertices1}
\end{align}
which includes the result of the previous section.}
The evaluation of the remaining vertex functions shown in Fig.~\ref{fig:loops}(e), however, results in two sets of equations
\begin{align}
 \left[\mathcal{S} \vec{\matrix{V}}_{(5)}^j\right]_{\beta\gamma}&=\frac{1}{2}\sigma_{\beta\rho} J_\sigma^{\rho\delta} F_{\sigma j}^\dagger(\omega)\sigma_{\delta\gamma},
\notag
\\
 \left[\mathcal{S}^\dagger \vec{\matrix{V}}_{(6)}^j\right]_{\beta\gamma}&=\frac{1}{2} J_\sigma^{\beta\gamma} F_{\sigma j}^\dagger(\omega),
 \label{eqn:loop_vertices2}
\end{align}
that also involve the adjoint of $\mathcal{S}$ defined by Eq.~(\ref{eqn:Sadjoint}), as derived explicitly in Appendix~\ref{app:vertexfunctions}. The sum of the five diagrams in Fig.~\ref{fig:loops} is then given by
\begin{align}
 \label{eqn:Q}
 \mathscr{L}(\omega) =\sum_{n=1}^5 \mathscr{L}_n(\omega) =\frac{1}{2}F_{i\alpha}(\omega) J_\alpha^{\beta\gamma}\left[\vec{\matrix{V}}^{j}(\omega)\right]_{\beta\gamma}+\hc,
\end{align}
where $\vec{\matrix{V}}^{j}(\omega)=\sum_{n=1}^5 \vec{\matrix{V}}^{j}_{(n)}(\omega)+\matrix{\sigma} \vec{\matrix{V}}_{(6)}^j(\omega)\matrix{\sigma}$.
In the following, we present explicit solutions to these equations for the cases of one-species and multi-species reaction networks.

\subsection{Explicit solution for one molecular species}

We consider the case of chemical reactions involving only a single species, in which the expressions derived in the previous section can readily be solved in closed form. For brevity, we set $D=D_{11}$, $J=J_1^1$, $D^{(1)}=D_{11}^{(1)}$, $J^{(1)}=J_{1}^{1(1)}$, $D_3=D_{111}$, and $\epsilon=\epsilon_1$. The power spectrum to order $\Omega^{-2}$ is then given by a sum of four contributions,
\begin{align}
\label{single:Spectrum}
 {S}(\omega)=\frac{1}{\Omega} \Delta(\omega) + \frac{1}{\Omega^2}\mathscr{T}(\omega)+ \frac{1}{\Omega^2}\mathscr{S}(\omega) + \frac{1}{\Omega^2}\mathscr{L}(\omega) + O(\Omega^{-3}).
\end{align}
The first term is simply the contribution from the LNA, which evaluates to
\begin{align}
\label{single:Delta}
\Delta(\omega)=\frac{D}{J^2+\omega^2} .
\end{align} 
The second term is given by 
\begin{align}
\label{single:TSum}
   &\mathscr{T}(\omega) = \frac{\epsilon}{J^2+\omega^2}
   \left[ 
D'  + 
J' 
\frac{J D}{J^2+\omega^2}  \right],
\end{align}
while the third reads
\begin{align}
\label{single:SSum}
   &\mathscr{S}(\omega) = \frac{1}{J^2+\omega^2}
   \left[ 
D^{(1)} +
D''\sigma +
(
    2{J^{(1)}} 
  + J''\sigma
)
\frac{J D}{J^2+\omega^2}  \right].
\end{align}
The remaining term is obtained by solving Eqs.~(\ref{eqn:loop_vertices1}) and (\ref{eqn:loop_vertices2}) using $\mathcal{S}^{-1}=(2J- i\omega)^{-1}$, and substituting the result in Eq.~(\ref{eqn:Q}). We find
\begin{align}
\label{single:QSum}
\mathscr{L}(\omega)
=
&\frac{1}{J^2+\omega^2}
\frac{J'}{(2J)^2+\omega^2}
\left[
D(2J^2-\omega^2)\frac{D'+2\sigma J'}{J^2+\omega^2}
+ 2J ( 2 D' + D_3 + \sigma^2 J')
\right],
\end{align}
which concludes our derivation. In Appendix \ref{app:VKex} we have verified the agreement of {the above} result with a particular case discussed by Calisto and Tirapegui \cite{calisto1993}.


\subsection{Explicit solution for general multi-species reaction networks}
\label{sec:explicit}

General expressions for the higher order corrections can be obtained via the diagonalization of the Jacobian $\matrix{J}$ of the macroscopic rate equations, which satisfies
\begin{align}
 \matrix{U}\matrix{J}\matrix{U}^{-1} = \text{diag}(\lambda_1,\ldots,\lambda_N).
\end{align}
Here, the matrix on the right hand side is diagonal, with $\lambda_1,\ldots,\lambda_N$ the eigenvalues of $\matrix{J}$. The spectral matrix $\matrix{S}$ can then be transformed as
\begin{align}
 \tilde{\matrix{S}}(\omega)=\matrix{U} \matrix{S}(\omega) \matrix{U}^T = 
 \frac{1}{\Omega}\tilde{\matrix{\Delta}}(\omega) + \frac{1}{\Omega^2}\tilde{\mathscr{T}}(\omega)+ \frac{1}{\Omega^2}\tilde{\mathscr{S}}(\omega) + \frac{1}{\Omega^2}\tilde{\mathscr{L}}(\omega) + O(\Omega^{-3}).
\end{align}
Applying the LNA, we find explicitly
\begin{align}
 \label{eqn:LNA_spectralMatrix}
 \tilde{{\Delta}}_{ij}(\omega)=\frac{\tilde{D}_{ij}}{(\lambda_i-i\omega)(\lambda_j+i\omega)}, 
\end{align}
where the coefficient is obtained from the transformed diffusion matrix $\tilde{\matrix{D}}$. The general transformation rules for the system size coefficients, Eq.~(\ref{eqn:SSEcoeffs}), are given by
\begin{align}
{\tilde{D}^{(n)}}_{ij..r} &=
U_{i\alpha} U_{j\beta} \ldots U_{r\rho}{D}^{(n)}_{\alpha\beta..\rho},\notag\\
{\tilde{J}{}^{(n)}}_{ij..r}^{st..z} &=
(U_{i\alpha}U_{j\beta} \ldots U_{r\rho}) (U^{-1}_{\sigma s} U^{-1}_{\tau t} \ldots U^{-1}_{\zeta z})
{{J}{}^{(n)}}_{\alpha\beta..\rho}^{\sigma \tau..\zeta}.
\end{align}
In order to evaluate these, we transform $\tilde{\mathcal{S}}=\matrix{U}\mathcal{S}\matrix{U}^T$ such that for any $\tilde{\matrix{V}}=\matrix{U}\matrix{V}\matrix{U}^T$, we have
$
 [\matrix{\tilde{S}}\matrix{\tilde{V}}]_{ij}= (\lambda_i+\lambda_j-i\omega)[\matrix{\tilde{V}}]_{ij},   
$
and, hence,
\begin{align}
 \label{eqn:diagS}
 [\matrix{\tilde{V}}]_{ij}=\frac{1}{\lambda_i+\lambda_j-i\omega} [\matrix{\tilde{M}}]_{ij}.
\end{align}
Using Eqs.~(\ref{eqn:T1}) through (\ref{eqn:T34}) and (\ref{eqn:T56}), we find
\begin{align}
\label{eqn:result_TSum}
   &\tilde{\mathscr{T}}_{ij}(\omega) = \frac{1}{2}\frac{\delta_{i\alpha}\delta_{j\beta}}{(\lambda_i-i\omega)(\lambda_j+i\omega)}
   \left[ 
\tilde{J}_{\alpha\beta}^\gamma 
+ 
 {\tilde{J}}_{\alpha}^{\mu\gamma}
\frac{\delta_{\mu\nu}{\tilde{D}}_{\nu\beta}}{\lambda_{\nu}-i\omega}  \right] \tilde{\epsilon}_\gamma 
+ (\omega \to -\omega, \alpha\leftrightarrow\beta)
 \end{align}
and
\begin{align}
\label{eqn:result_SSum}
   &\tilde{\mathscr{S}}_{ij}(\omega) = \frac{1}{2}\frac{\delta_{i\alpha}\delta_{j\beta}}{(\lambda_i-i\omega)(\lambda_j+i\omega)}
   \left[ 
\tilde{D}_{\alpha\beta}^{(1)} + 
{\tilde{J}}_{\alpha\beta}^{\gamma\delta}\tilde\sigma_{\gamma\delta}+
(
    2{\tilde{J}{}^{(1)}}_\alpha^\mu  + 
    {\tilde{J}}_{\alpha}^{\mu\gamma\delta}\tilde\sigma_{\gamma\delta}
)
\frac{\delta_{\mu\nu}{\tilde{D}}_{\nu\beta}}{\lambda_{\nu}-i\omega}  \right] 
+ (\omega \to -\omega, \alpha\leftrightarrow\beta).
 \end{align}
Further, solving Eq.~(\ref{eqn:loop_vertices1}) and (\ref{eqn:loop_vertices2}) using Eq.~(\ref{eqn:diagS}) and substituting the result in Eq.~(\ref{eqn:Q}), we deduce
\begin{align}
\label{eqn:QSum1}
\tilde{\mathscr{L}}_{ij}(\omega) = 
&\frac{1}{2}\frac{\delta_{i\alpha}\delta_{j\beta}}{(\lambda_i-i\omega)(\lambda_j+i\omega)}
\frac{\tilde{J}_\alpha^{\theta\zeta} \delta_{\mu\theta}\delta_{\gamma\zeta}}{\lambda_\mu+\lambda_\gamma-i\omega}
\times \notag\\
&\left.
\left[
(\tilde{J}_{\mu\gamma}^{\delta}
+2\tilde\sigma_{\mu\rho}\tilde{J}_{\gamma}^{\rho\delta}) \delta_{\delta l} \frac{\tilde{D}_{l\beta}}{\lambda_l-i\omega}
+ 2 \tilde{J}_{\mu\beta}^\gamma
+ \tilde{D}_{\mu\gamma\beta} + \tilde\sigma_{\mu\rho} \tilde{J}_\beta^{\rho\delta}\tilde\sigma_{\delta\gamma}
\right]
\right. 
+ (\omega \to -\omega, \alpha\leftrightarrow\beta),
\end{align}
where we have used the symmetry of the last summand in the angled brackets in Eq.~(\ref{eqn:QSum1}). Note also that these equations are to be summed over all Greek indices, according to our summation convention. 

{From the analytical form of these correction terms, we recover the well-known fact that they vanish for reaction networks with at most unimolecular reactions; see for instance \cite{warren2006}. {Note that} the microscopic rate functions in Eq.~(\ref{eqn:microscopicRates}) are at most linear in the concentrations, and independent of the system size $\Omega$, {in that case}. The former {observation shows} that all vertices with multiple upper indices are identically zero, while the latter implies that $\tilde{D}_{\alpha\beta}^{(1)}$ and ${\tilde{J}{}^{(1)}}_\alpha^\mu$ vanish {and, hence, that} the corrections given by Eqs.~(\ref{eqn:result_SSum}) and (\ref{eqn:QSum1}) are absent. It also follows {from} Eq.~(\ref{eqn:EMRE}) that (\ref{eqn:result_TSum}) vanishes, as the mean concentrations predicted by the CME agree to this order with the prediction from the rate equations. These correction terms hence stem from the nonlinearity in the law of mass action, and {thus} must increase with the size of the rate constants in the bimolecular reactions.
}

{It can also be deduced from Eqs.~(\ref{eqn:result_TSum}) through (\ref{eqn:QSum1}) that the nonlinear corrections to the power spectrum involve denominators $(\lambda_i-i\omega)(\lambda_j+i\omega)$ {and, hence, that they} contribute to the amplitude of the power spectrum at the frequencies of the linear relaxation modes. Further, Eq.~(\ref{eqn:QSum1}) involves the denominator $\lambda_\mu+\lambda_\gamma-i\omega$ which stems from nonlinear relaxation modes that are not described by the linear response. These additional frequencies appear {in} the sums of pairs of the eigenvalues of the linear relaxation modes. {Such} corrections are particularly important when close to bifurcation points; for example, in the case of noise-induced oscillations below a Hopf bifurcation point, {they} lead to a secondary harmonic, and a zero-frequency component in the power spectrum. {This phenomenon has been reported} by some of {the} authors in \cite{thomas2013}, {where the Chemical Langevin Equation was used}, and is derived here explicitly from the CME for the first time.
}

\section{Applications}

\subsection{Dimerization}
As a first example, we consider bursty expression of a protein and dimerization. Bursty synthesis may arise through the finite lifetimes of the associated transcripts \cite{shahrezaei2008}. A simple reaction system describing synthesis in bursts of size $m$ can be formulated as
\begin{align}
\varnothing \xrightarrow{k_\text{in}} m X , \ \ 2 X \xrightarrow{1} \varnothing
\end{align}
The expansion coefficients can be calculated from the stoichiometric matrix $\matrix{S}=(m,-2)$ and the microscopic rate functions, which are given by the series
\begin{align}
\hat{\vec{f}}({n},\Omega) = {\vec{f}}\left(\frac{{n}}{\Omega}\right) + \frac{1}{\Omega} {\vec{f}}^{(1)}\left(\frac{{n}}{\Omega}\right)
\end{align}
with $\vec{f}(\phi)=(k_\text{in},\phi^2)^T$ and $\vec{f}^{(1)}(\phi)=(0,-\phi)^T$, from which one {deduces} $J=-4\phi$ and $D=2(2+m)\phi^2$. The concentration predicted by the macroscopic rate equation, Eq.~(\ref{eqn:REs}), is obtained as $\phi=(m k_\text{in}/2)^{1/2}$.
Hence, the power spectrum and variance, as found from the LNA, reads
\begin{align}
 \label{dim:delta}
 \Delta(\omega)=\frac{2 (2+m) \phi ^2}{16 \phi ^2+\omega ^2}, \ \
 \sigma=\frac{2+m}{4}\phi.
\end{align}
Further, using $J'=-4$ and $D^{(1)}=-4\phi$ the mean concentration is calculated from Eq.~(\ref{eqn:EMRE2}) as
\begin{align}
 \left\langle\frac{n}{\Omega}\right\rangle = \phi + \frac{1}{\Omega}\frac{2-m}{8} + O(\Omega^{-2}).
\end{align}
The higher order correction to the intrinsic noise power spectrum can now be derived by substituting the coefficients $J''=0$, $D'=4(2+m)\phi$, $D''=4(2+m)$ $J^{(1)}=2$, and $D_3=2 \left(m^2-4\right)\phi^2$ into Eqs.~(\ref{dim:delta}), (\ref{single:TSum}) and (\ref{single:QSum}), and {by making use of} Eq.~(\ref{single:Spectrum}). The result is
\begin{align}
 S(\omega)=\frac{1}{\Omega}\Delta(\omega)+ \Omega ^2\left[
\frac{64 (2+m) \phi ^3 \left(2 (2+m) \phi ^2-(m-1) \omega ^2\right)}{\left(16 \phi ^2+\omega ^2\right)^2 \left(64 \phi ^2+\omega ^2\right)}\right].
\end{align}
{
Finally, the Fourier transform of the above expression yields the autocorrelation function
\begin{align}
\label{eqn:dimer_sigma}
&\left\langle 
\left(\frac{n(t+\tau)}{\Omega} - \left\langle\frac{n}{\Omega}\right\rangle\right) 
\left(\frac{n(t)}{\Omega} - \left\langle\frac{n}{\Omega}\right\rangle\right) 
\right\rangle
=\int \frac{\dd \omega}{2\pi}{\rm e}^{i\omega\tau} S(\omega)
\notag\\
&=
  {\rm e}^{-4\phi\tau} \left( \frac{\sigma}{\Omega}+
\frac{1}{\Omega^2}
\frac{(2+m)}{96 }(14-13m +12(3m-2)\tau\phi )\right) 
+
\frac{1}{\Omega^2}
\frac{(2+m)}{96 }{\rm e}^{-8 \phi\tau}\left(11 m-10 \right) + O(\Omega^{-3}).
\end{align}
The first term includes a $\Omega^{-2}$ correction to the amplitude of the linear response, while the second represents additional relaxation terms that decay twice as fast; see also the discussion concluding Section \ref{sec:explicit}.}

{
In order to verify the accuracy of our analysis, we calculate the $\Omega^{-2}$ correction to the variance given by the LNA using Eq.~(25) together with Eqs.~(28) through (30) from \cite{grima2011}: 
\begin{align}
 \left\langle 
\left(\frac{n(t)}{\Omega} - \left\langle\frac{n}{\Omega}\right\rangle\right)^2
\right\rangle=\frac{1}{\Omega}\sigma+\frac{1}{\Omega^2} \frac{(2+m)(2-m)}{48} + O(\Omega^{-3}),
\end{align}
which agrees with Eq.~(\ref{eqn:dimer_sigma}) {when} evaluated at $\tau=0$.
The particular case of $m=2$, which obeys detailed balance with a Poissonian steady state, has been considered by Chaturvedi and Gardiner \cite{chaturvedi1978} (with $\kappa_2=2$ in their notation). They obtained the autocorrelation function
\begin{align}
\frac{1}{\Omega} {\rm e}^{-4\phi\tau} \sigma +
\frac{1}{\Omega^2} \frac{{\rm e}^{-8 \tau  \phi } \left(1+{\rm e}^{4 \tau  \phi } (4 \tau  \phi -1)\right)}{2}+ O(\Omega^{-3}),
\end{align}
which agrees with Eq.~(\ref{eqn:dimer_sigma}) in this case. It is interesting to observe that, despite the detailed balance of the reactions, the nonlinearity in the law of mass action manifests itself in the non-trivial dependence of the $\Omega^{-2}$ terms.
}
\subsection{Noise-induced oscillations in the Brusselator reaction}

\begin{figure}[t]
   \centering
   \includegraphics[width=1\textwidth]{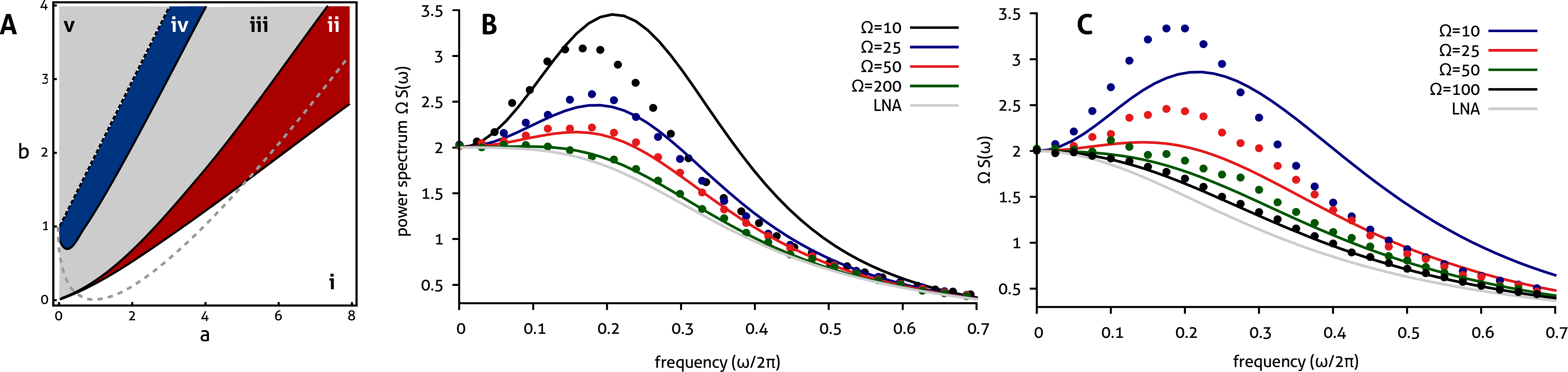}
   \caption{\textbf{Noise-induced oscillations in the Brusselator reaction:}
   (\textbf{A}) {Comparison of the} parametric dependence of noise-induced oscillations for species $X$ {with} the prediction {from} the LNA and {with that obtained by} including higher order corrections in the system size expansion, for $\Omega=20$. In region (i, white area), both theories predict that no oscillations will occur; in (ii, red), the higher order {expansion} predicts noise-induced oscillations which are missed by the {LNA}. Note that these oscillations appear both for deterministically stable nodes and {foci} (separated by {a dashed gray} line). In regions (iii, gray) and (iv, blue), both theories predict noise-induced oscillations. Region (iv, blue) includes the appearance of secondary harmonics {that are} not captured by the LNA close to the deterministic Hopf bifurcation ({dashed black} line). {The deterministically oscillatory regime in (v, gray) is not described by either theory.} 
   (\textbf{B}) The intrinsic noise power spectrum in the case of a deterministic {focus} ($k_0=1=k_1$, $a=4.5$, $b=2$) is shown for decreasing values of $\Omega$. {For large $\Omega$, the monotonous dependence on frequency is captured by the LNA (dashed gray line).} {A decrease in volume reveals noise-induced oscillations that are well described} by the $\Omega^{-2}$ corrections {(blue and red lines)}, but not by the LNA. These predictions are verified by stochastic simulation (dotted). Reducing the volume to $\Omega=10$, we observe that our theory only qualitatively accounts for the frequency dependence. 
   (\textbf{C}) A similar dependence is observed in the parameter regime {corresponding to a} deterministic node ($k_0=1=k_1$, $a=7.5$, $b=3$). {However,} for small volumes, the higher order {expansion only yields a qualitative description for} the power spectrum of the noise-induced oscillations. (Note that in (A) and (B), the spectrum has been multiplied by $\Omega$ such that the prediction from the LNA is the same for all system sizes.)
}
\label{fig:brussel}
\end{figure}
The Brusselator describes a commonly studied set of autocatalytic oscillatory reactions \cite{prigogine1968}, which involve two species interacting via
\begin{align}
\varnothing \xrightleftharpoons[k_1]{k_0} X, \ \ X\xrightarrow{b} Y , \ \ 
2X + Y\xrightarrow{a} 3X.
\end{align}
The form of the propensities $\Omega \hat{\vec{f}}(\vec{n},\Omega)=(\Omega k_0, k_1 n_X, b n_X, a n_Y\, n_X(n_X-1)\Omega^{-2})$ implies the following expansion of the rate functions, Eq.~(\ref{eqn:microscopicRates}),
\begin{align}
\hat{\vec{f}}(\vec{n},\Omega) = {\vec{f}}\left(\frac{\vec{n}}{\Omega}\right) + \frac{1}{\Omega} {\vec{f}}^{(1)}\left(\frac{\vec{n}}{\Omega}\right),
\end{align}
where ${\vec{f}}(\vec{\phi})=(k_0, k_1 [X], b [X], a [Y][X]^2 )$ and ${\vec{f}}^{(1)}(\vec{\phi})=(0, 0, 0, -a[Y][X] )$. 
This expansion, together with the stoichiometry
\begin{align}
 \matrix{R}= 
\left(
\begin{array}{cccc}
 1 & 1 & -1 & 1 \\
 0  & 0 & 1 & -1 
\end{array}
\right),
\end{align}
determines the system size coefficients, Eq.~(\ref{eqn:SSEcoeffs}). Specifically, the solution of the macroscopic rate equations, Eq.~(\ref{eqn:REs}), is given by $[X]=\frac{k_0}{k_1}$ and $[Y]= \frac{b k_1}{a k_0}$. 
The Jacobian of {these equations then reads}
\begin{align}
 \label{brussel:Jac}
 \matrix{J}=\left(
\begin{array}{cc}
 b-k_1 & a[X]^2 \\
 -b    & -a[X]^2 \\
\end{array}
\right).
\end{align}
Using Eq.~(\ref{eqn:EMRE2}), the mean concentrations are obtained as
\begin{align}
 \left\langle \frac{n_X}{\Omega} \right\rangle &= [X] + O(\Omega^{-2}), \notag\\
 \left\langle \frac{n_Y}{\Omega} \right\rangle &= [Y] + \frac{1}{\Omega} 
 \frac{2 b \left(2 a [X]^2-b\right)}{a [X]^2 \left(a [X]^2+k_1-b\right)} + O(\Omega^{-2}),
\end{align}
which are accurate to order $\Omega^{-1}$. Hence, to this order, the prediction from the CME agrees with the one {from} the macroscopic rate equations for species $X$, but not for species $Y$. The leading order corrections to the power spectrum can be derived either algebraically or by making use of the eigenvalue representation, as follows. The eigenvalues of the Jacobian in Eq.~(\ref{brussel:Jac}) are given by
\begin{align}
 \label{brussel:eigenvals}
 \lambda_{1,2} =
-\frac{1}{2}\left(a[X]^2+k_1-b\right)
\pm\frac{1}{2} \sqrt{\left(a[X]^2+k_1-b\right)^2-{4 a k_0}[X]},
\end{align}
while the corresponding eigenvectors are
$U_{i1} = b,$ 
and 
$U_{i2} = (\lambda_i + k_1-b)$, for $i=1,2$. In particular, we see that the deterministic system undergoes a Hopf bifurcation when the real part of the above eigenvalues becomes positive. Since, strictly speaking, the system size expansion only applies to monostable systems, we require $b<a[X]^2+k_1$. The resulting analytical expressions for the power spectrum are, however, rather involved. Hence, we restrict our analysis to the study of species $X$ in the case where $k_0=1=k_1$. On the basis of the LNA, we obtain
\begin{align}
 \Delta_X(\omega)=\frac{2 \left(a^2+(b+1) \omega ^2\right)}{a^2+\omega ^2 \left((a-b)^2+1-2 b\right)+\omega ^4}.
\end{align}
The intrinsic noise power spectrum, including corrections of order $O(\Omega^{-2})$, is then obtained via the procedure derived in Section 4 and reads
\begin{align}
 \label{brussel:spec}
 S_X(\omega) =  \frac{1}{\Omega}\Delta_X(\omega)+ \frac{1}{\Omega^2} \frac{\sum_{i=1}^5 a_i\omega^{2i}}{A(\omega)} + O(\Omega^{-3}),
\end{align}
where the denominator is given by
\begin{align}
 A(\omega) =
 & (a-b+1) \left((a-b+1)^2+\omega ^2\right) \times \notag\\
 & \left(a^2+\omega ^2 \left((a-b)^2-2 b+1\right)+\omega ^4\right)^2
\left(16 a^2+4 \omega ^2 \left((a-b)^2-2 b+1\right)+\omega ^4\right).
\end{align}
Here, the coefficients $a_i$ are defined as
\begin{align}
a_1= & 32 a^4 b 
 \left(-3 (11 a+9) b^3+((58 a+75) a+42) b^2-(((49 a+96) a+72) a+19) b+4 ((a+3) (4 a+3) a+3) a+8 b^4\right), \notag
\\
a_2= & 8 a^2 b \left(16 a^6+36 a^5 (3-2 b)+4 a^4 ((34 b-69) b+17)
\right. \notag \\
&
\left.
+a^3 ((2 (121-72 b) b-359) b+40)+a^2 (3 b (b ((32 b-37) b+131)-12)-172)
\right. \notag \\
&
\left.
+2 a (b (((10 (3-2 b) b-133) b+88) b+63)-18)+(b-1)^2 (((8 b-7) b+82) b+1) b\right)
, \notag \\
a_3= & 8 b \left(12 a^7+a^6 (8-40 b)+a^5 \left(32 b^2-58 b+3\right)+a^4 \left(37 b^3+128 b^2+b-55\right)+a^3 (63-2 b ((b (44 b+59)-52) b+24))
\right. \notag \\
&
\left.
+a^2 \left(b \left(b \left(\left(70 b^2+34 b-189\right) b+228\right)-159\right)-52\right)-4 a (b-1)^2 (((7 b-4) (b+2) b+3) b+3)+5 (b-1)^4 (b+1)^2 b\right)
\notag,\\
a_4= & 8 b \left(-a^5+3 a^4 (b-1)+a^3 (14-b (4 b+5))+a^2 (b (b (7 b+59)-75)-17)
\right. \notag\\
&
\left.
+a (((42-b (9 b+35)) b+21) b+17)+4 \left(b^2-1\right)^2 b\right),
\notag \\
a_5= & -8 b \left(a^3+a^2 (7 b+1)-a (7 (b+1) b+5)+(b+1)^2 b\right).
\end{align}

The above explicit expression for the power spectrum can be used to quantify the periodicity of typical trajectories of the underlying stochastic process, which is for instance of importance in the study of single-cell oscillations \cite{westermark2009}. Specifically, if the deterministic stability is given by a stable fixed point, a peak in the intrinsic noise power spectrum indicates a noise-induced oscillation. The parametric dependence of these oscillations is analyzed in Fig.~\ref{fig:brussel} using Eq.~(\ref{brussel:spec}) for a system size of $\Omega=20$. We find that the LNA predicts the fluctuations to decay monotonically with increasing frequency in parameter regions (i) and (ii), while it postulates noise-induced oscillations in parameter regions (iii) and (iv). By contrast, taking into account higher order corrections as given by the second term in Eq.~(\ref{brussel:spec}), we infer that the dynamics is also oscillatory in region (ii), which is at odds with the prediction from the LNA. In Figs.~\ref{fig:brussel}B and C, we verify this dependence {in two parameter regimes}, one in which the eigenvalues of the Jacobian, as given in Eq.~(\ref{brussel:eigenvals}), are complex, and another in which these eigenvalues are purely real, corresponding to stable node oscillations \cite{toner2013}. There, we also show the dependence of the oscillations on the system size $\Omega$. {Finally,} we note that our theory predicts the presence of secondary harmonics in the vicinity of the Hopf bifurcation (dashed black) that cannot be described by the LNA. An approximate treatment that accounts quantitatively for these effects has been given by some of the authors elsewhere \cite{thomas2013}.

\FloatBarrier

\section{Discussion}

In this article, we have presented a general diagrammatic approach for the systematic evaluation of multi-time correlation functions from the system size expansion that goes beyond the common LNA. 
As we have demonstrated, application of the proposed {set of Feynman rules allows for the calculation of higher order corrections in the expansion to be performed in an algorithmic fashion}. In brief, our methodology {permits the representation of the corresponding} correlation functions by a sum of diagrams which can be truncated to the desired order in the inverse system size. Each diagram is itself derived from a set of Feynman rules and can be readily translated into certain integrals in the frequency domain. The technique therefore represents an effective bookkeeping device for the purposes of detailed calculation {on the basis of} the system size expansion.

{We emphasize that our approximation for the generating functional does not exceed the accuracy provided by the LNA, from which correlation functions are ultimately generated. Renormalization theory typically employs the Legendre transform of the cumulant generating functional for that purpose; the latter can be constructed either from a diagrammatic expansion of the associated correlation functions, or of the vertex functions alone \cite{srednicki,tauber2014,zinn2002}.}

{Different diagrammatic methods have been applied to obtain stochastic perturbation series using the Doi-Peliti coherent state path-integral in homogeneous or spatial models of population dynamics \cite{peliti1985,moloney2006,tauber2012,tauber2014}. The nonlinear reactions are therein treated as a perturbation to a functional of non-interacting, or ``free'', species. In the present {perturbative ansatz}, however, the Gaussian functional $Z_0$ takes the role of a ``free theory'' which arises from the exact generating functional via the LNA in the limit of large system size, i.e., via a linearization of the concentration fluctuations about the macroscopic limit of the full interacting reaction network. As we have demonstrated, {our methodology} is significantly simpler than present techniques for the system size expansion \cite{grima2011,cianci2012,scott2012,thomas2013} that are equivalent, but {that} rely instead on the explicit solution of high-dimensional systems of ordinary differential equations. In particular, it does not involve the calculation of higher moments. Similar perturbation expansions may as well be obtained, for instance, through the functional integral of the Doi-Peliti approach or the Poisson representation \cite{droz1994}. Our choice of the equivalent Calisto-Tirapegui path-integral enjoys the particular advantage that the form of the expansion is in one-to-one correspondence with the system size expansion, as originally proposed by van Kampen.}

We have demonstrated the resolution of the diagrammatic expansion to order $\Omega^{-2}$ by deriving closed-form expressions for the intrinsic power spectra of general biochemical networks. The {resulting} expressions are more accurate than the ones obtained from the common LNA. We have illustrated the {utility of our approach} through the analytical computation of the autocorrelation function in a dimerization reaction and {through a study} of noise-induced oscillations in the Brusselator reaction. Specifically, we have demonstrated that our approach allows {for the identification of} noise-induced oscillations in parameter regimes that are not captured by the conventional LNA, including {in} regimes in which a deterministic stability analysis predicts either stable foci or nodes. These discrepancies are of particular interest for the analysis of single cell rhythms that are sustained by noise due to low number of molecules \cite{westermark2009,toner2013,thomas2013}. Our approach may hence serve to deepen our understanding of stochastic dynamics in biochemical networks.

\section*{Acknowledgments}
We thank Claudia Cianci and Pierce Munnelly for interesting discussions. RG gratefully acknowledges support from SULSA (Scottish Universities Life Sciences Alliance). 

\appendix

\renewcommand{\theequation}{A\arabic{equation}}    
\setcounter{equation}{0}  

\section{Appendix}

\subsection{The functional integral representation}
\label{app:fi}

{
{Here, we} derive the functional integral representation of the CME
\begin{align}
 \label{eqn:transitionP}
 \frac{\partial}{\partial t} P(\vec{n},t|\vec{n}',t') &= \Omega\mathcal{L}(-\nabla_n,\vec{n}) P(\vec{n},t|\vec{n}',t'),
\end{align}
where $\vec{n}$ {denotes} the $N$-dimensional state vector and the operator $\mathcal{L}(-\nabla_n,\vec{n})=\sum_j({\rm e}^{(-\nabla^T \matrix{R})_j}-1)\hat{f}_j\left(\frac{\vec{n}}{\Omega},\Omega\right)$ is the transition matrix of the the CME, Eq.~(\ref{eqn:CME}), that  is ``normally ordered" in the sense that it has all derivatives $\nabla_n$ to the left and all $\vec{n}$ to the right.
}

\subsubsection*{Operator formulation of the {CME}}

In a first step, we determine a set of eigenfunctions for the following two operators
\begin{align}
 \hat{\vec{Q}} \equiv \frac{\vec{n}}{\Omega}, \ \ \hat{\vec{P}}\equiv-\nabla_n^T,
\end{align}
which satisfy the commutation relation $[\hat{\vec{Q}},\hat{\vec{P}}]=\Omega^{-1}\matrix{1}$. The {normal ordering} of $\mathcal{L}$ suggests that a suitably chosen set will satisfy the eigenvalue equations
$\hat{\vec{Q}}|\vec{Q}\rangle=\vec{Q}|\vec{Q}\rangle$ and $\langle \vec{P}|\hat{\vec{P}}=\langle \vec{P}|\vec{P}$.
Using the identity ${\rm e}^{ -\Omega \hat{\vec{P}}\vec{Q}} \hat{\vec{Q}} {\rm e}^{\Omega\hat{\vec{P}}\vec{Q}}=\hat{\vec{Q}} +\vec{Q}$ \cite{orszag2008} together with $\hat{\vec{Q}}|0\rangle=0$, we verify that 
\begin{align}
|\vec{Q}\rangle\equiv| Q_0\rangle \oplus| Q_1\rangle \oplus\dots | Q_N\rangle =\exp(\Omega\hat{\vec{P}}\vec{Q})|0\rangle,
\end{align}
is an eigenvector of $\hat{\vec{Q}}$ with eigenvalue $\vec{Q}$. Defining 
\begin{align}
\langle \vec{P} | = \int \dd {\vec{Q}} {\rm e}^{\Omega \vec{P}\vec{Q}} \langle \vec{Q} |,
\end{align}
we verify that the above is an eigenvector of $\hat{\vec{P}}$ with eigenvalue $\vec{P}$. Likewise, we have
\begin{align}
\langle \vec{Q} | = \Omega^N\,\int  \frac{\dd\vec{P}}{2\pi i} {\rm e}^{-\Omega \vec{P}\vec{Q}} \langle \vec{P} |, 
\end{align}
where the domain of $\vec{P}$ is the imaginary axis such that $\langle \vec{Q}| \vec{Q}' \rangle=\delta(\vec{Q}- \vec{Q}')$.
Note also that these states are not orthogonal, as
\begin{align}
\langle \vec{P}|\vec{Q}\rangle={\rm e}^{\Omega\vec{P}\vec{Q}}, \ \ \langle \vec{Q}| \vec{P}\rangle={\rm e}^{-\Omega\vec{P}\vec{Q}}.
\end{align}
The partition of unity can be derived as
\begin{align}
\label{eqn:completeness}
\underbar{1}=\Omega^{N}\int \frac{\dd \vec{Q} \dd \vec{P}}{2\pi i } {\rm e}^{-\Omega\vec{P}\vec{Q}} |\vec{Q} \rangle\langle \vec{P}|,
\end{align}
which follows from
\begin{align}
\label{eqn:unityproof}
\Omega^{N}\int\frac{ \dd \vec{Q} \dd \vec{P}}{2\pi i} {\rm e}^{-\Omega\vec{P}\vec{Q}} |\vec{Q} \rangle\langle \vec{P}|
&= \Omega^{N}\int \dd {\vec{Q}} \int \dd \hat{\vec{Q}'} \int \frac{\dd \tilde{\vec{P}}}{2\pi} {\rm e}^{i\Omega\tilde{\vec{P}}({\vec{Q}}'-\vec{Q})} |\vec{Q} \rangle \langle \vec{Q}'|    \notag\\
&= \int \dd {\vec{Q}}  \int \dd {\vec{Q}'} \delta({\vec{Q}}'-\vec{Q}) |\vec{Q} \rangle \langle \vec{Q}'| \notag\\
&= \int \dd {\vec{Q}} |\vec{Q} \rangle \langle \vec{Q}|.
\end{align}
Here, we have changed the integration variable via $\vec{P}\to i\tilde{\vec{P}}$ in the first line. Similarly, we find $\underbar{1}=\Omega^N \int \frac{\dd\vec{P}}{2\pi i} |\vec{P}\rangle\langle \vec{P}|$.

{
We now associate an abstract state vector $|\Psi(t)\rangle$ such that $\langle \vec{Q} | \Psi(t) \rangle=P(\vec{Q},t)$ is the probability density function to observe the concentrations $\vec{Q}$ at time $t$. Using the resolution of unity, Eq.~(\ref{eqn:completeness}), we can write
\begin{align}
 |\Psi(t)\rangle = \int \dd \vec{Q}  |\vec{Q}\rangle P(\vec{Q},t).
\end{align}
Multiplying Eq.~(\ref{eqn:transitionP}) with $|\vec{Q}\rangle$ and integrating over all $\vec{Q}$, we find that $|\Psi\rangle$ satisfies the evolution equation
\begin{align}
 \label{eqn:SchroedingerTypeEquation}
 \frac{\partial}{\partial t}|\Psi\rangle = \Omega\mathcal{L}(\hat{\vec{P}},\hat{\vec{Q}})|\Psi\rangle.
\end{align}
The above result is exactly equivalent to the Fock space representation of Doi-Peliti \cite{peliti1985,kamenev2011,tauber2014}, {which uses} creation-annihilation operators defined by $a_i^\dagger|\vec{n}\rangle=|n_1,\ldots,n_i+1,\ldots,n_N\rangle$ and $a_i|\vec{n}\rangle=n_i|n_1,\ldots,n_i-1,\ldots,n_N\rangle$, as well as the Cole-Hopf transformation $a_i^\dagger={\rm e}^{\hat{P}_i}$ and $a_i={\rm e}^{\hat{P}_i}\Omega\hat{Q}_i$ such that $a_i^\dagger {a}_i=\Omega\hat{Q}_i$ \cite{kamenev2011}. For our purposes, the present parametrization of $\hat{\vec{Q}}$ and $\hat{\vec{P}}$ is, however, preferable, since it enables us to construct the functional integral for the molecular concentrations $\hat{\vec{Q}}$ {in a straightforward fashion}.}

Note also that by multiplying Eq.~(\ref{eqn:SchroedingerTypeEquation}) with the eigenstate $\langle 0 | = \int \dd \vec{Q} \langle \vec{Q} |$ belonging to $\vec{P}=0$, we must have
\begin{align}
 \frac{\partial}{\partial t}  \int \dd \vec{Q} P(\vec{Q},t) 
= \Omega\langle 0| \mathcal{L}(0,\hat{\vec{Q}}) | \Psi \rangle  \overset{!}{=} 0.
\end{align}
Hence, we require $\mathcal{L}(0,\hat{\vec{Q}})=0$ by conservation of probability, which is ensured by the {definition} of $\mathcal{L}$ given after Eq.~(\ref{eqn:transitionP}).

\subsubsection*{Path-integral formulation of the transition probability}

{The solution of Eq.~(\ref{eqn:SchroedingerTypeEquation}) for an ensemble initially prepared in state $|\vec{Q}'\rangle$ at time $t'$ is given by $|\Psi(t)\rangle={\rm e}^{\Omega\mathcal{L}(t-t')} |\vec{Q}' \rangle$. {Hence,} the transition probability $\langle\vec{Q}|\Psi(t)\rangle$ {reads}}
\begin{align}
\label{eqn:transprob}
P(\vec{Q},t|\vec{Q}',t')
&=\langle \vec{Q}| {\rm e}^{\Omega\mathcal{L}(t-t')} |\vec{Q}' \rangle\notag\\
&=\Omega^N \int \frac{\dd \vec{P}}{2\pi i}   {\rm e}^{-\Omega\vec{P}\vec{Q}}\langle \vec{P}| {\rm e}^{\Omega\mathcal{L}(t-t')}|\vec{Q}'\rangle.
\end{align}
Note that in the second line, we have made use of the relation in (\ref{eqn:completeness}).
We continue by slicing the time evolution from $t'$ to $t$ into $L-1$ pieces with a time step of $\delta t=(t-t')/L$, inserting Eq.~(\ref{eqn:completeness}) $L-1$ times into the above expression: 
\begin{align}
\langle \vec{P}|\, {\rm e}^{\Omega\mathcal{L}(t-t')}|\vec{Q}'\rangle=
\left(\prod_{k=1}^{L-1} \Omega^{N}\int\frac{\dd \vec{Q}_k\dd \vec{P}_k}{2\pi i} {\rm e}^{-\Omega\vec{P}_k \vec{Q}_k} \right)
\langle \vec{P}|\,{\rm e}^{ \delta t\Omega\mathcal{L}}|\vec{Q}_{L-1}\rangle
\langle \vec{P}_{L-1}|\,{\rm e}^{ \delta t\Omega\mathcal{L}}|\vec{Q}_{L-2}\rangle
\dots\langle \vec{P}_1|\,{\rm e}^{ \delta t\Omega\mathcal{L}}|\vec{Q}'\rangle.
\end{align}
Since $\mathcal{L}(\hat{\vec{P}},\hat{\vec{Q}})$ is normally ordered, we have 
\begin{align}
\langle\vec{P}_k|\mathcal{L}(\hat{\vec{P}},\hat{\vec{Q}})|\vec{Q}_{k-1}\rangle=\mathcal{L}(\vec{P}_k,\vec{Q}_{k-1})\langle\vec{P}_k|\vec{Q}_{k-1}\rangle.
\end{align}
The time-sliced transitions can then be evaluated in the limit as $L\to\infty$ when the time step $\delta t$ becomes infinitesimally small, as follows:
\begin{align}
\langle \vec{P}_k|{\rm e}^{ \delta t\mathcal{L}}|\vec{Q}_{k-1}\rangle&=\langle \vec{P}_k|\vec{Q}_{k-1}\rangle + \delta t\langle \vec{P}_k|\Omega\mathcal{L}(\hat{\vec{P}},\hat{\vec{Q}})|\vec{Q}_{k-1}\rangle+O(\delta t^2)\notag\\&\approx\exp\left(\Omega\vec{P}_k\vec{Q}_{k-1} +\delta t\Omega\mathcal{L}(\vec{P}_k,\vec{Q}_{k-1})\right),
\end{align}
which is correct up to $O(\delta t^2)$. Substituting back into Eq.~(\ref{eqn:transprob}), we find
\begin{align}
\label{eqn:pathintegral}
P(\vec{Q},t|\vec{Q}',t')= \Omega^N\lim\limits_{L\to\infty}
\left(\prod_{k=1}^{L-1}\int{\dd \vec{Q}_k} \right) \left(\prod_{k=0}^{L} \int\frac{  \dd \vec{P}_k}{2\pi i} \right)
\exp\left[-\sum_{k=0}^L \left( \Omega\vec{P}_k(\vec{Q}_{k}-\vec{Q}_{k-1}) -\delta t\Omega\mathcal{L}(\vec{P}_k,\vec{Q}_{k-1}) \right)\right],
\end{align}
which expresses the transition probability as a sum over paths of the stochastic process. Note that the factor $\Omega^N$ is the Jacobian determinant relating $P(\vec{Q},t|\vec{Q}',t')$ and $P(\vec{n},t|\vec{n}',t')$ Note also that the endpoints are not contained in the integration range {here}. 

\subsubsection*{Functional integral representation}
In order to calculate the generating functional given by Eq.~(\ref{eqn:CMEpathintegral}) in the main text, we replace the transition matrix in Eq.~(\ref{eqn:pathintegral}) by $\Omega\mathcal{L}\to\Omega\mathcal{L}+\hat{\vec{P}}\vec{j}^\star  +\vec{j}^T\hat{\vec{Q}}$. {Then, we} multiply the transition probability by the initial condition $\delta(\vec{Q}(t')-\vec{\phi}(t'))$, and integrate out the remaining state vectors. The result is the generating functional
\begin{align}
\label{app:GF}
Z(\vec{j},\vec{j}^\star)
&=\lim\limits_{L\to\infty}
\left(\prod_{k=0}^{L} \Omega^N \int\frac{ \dd \vec{Q}_k\dd \vec{P}_k}{2\pi i} \right) \times\notag\\
&\exp\delta t \sum_{k=0}^L \left(- \Omega\vec{P}_k\frac{(\vec{Q}_{k}-\vec{Q}_{k-1})}{\delta t} +\Omega\mathcal{L}(\vec{P}_k,\vec{Q}_{k-1})+ \vec{P}_k\vec{j}^\star  +\vec{j}^T\vec{Q}_{k-1} \right)
 \delta(\vec{Q}(t')-\vec{\phi}(t'))
 \notag\\
&\equiv \int \mathcal{D}\vec{P} \mathcal{D}\vec{Q} \exp \int_{t'}^{t} \dd t \left( -\Omega\vec{P}\dot{\vec{Q}} +\Omega\mathcal{L}(\vec{P},\vec{Q})+\vec{P}\vec{j}^\star +\vec{j}^T\vec{Q} \ \right)\delta(\vec{Q}(t')-\vec{\phi}(t')),
\end{align}
where the measure $\mathcal{D}\vec{P} \mathcal{D}\vec{Q}$ denotes the integral obtained by the limiting process in Eq.~(\ref{app:GF}).

\subsection{Sylvester matrix equation}
\label{app:sylvester}

We consider the following Fourier integral
\begin{align}
 \vec{\matrix{{V}}}(\omega)=\int_0^\infty {\dd \tau} {\rm e}^{(\matrix{J}-i\omega) \tau} \matrix{M}(\omega) {\rm e}^{\matrix{J}^T \tau},
\end{align}
which determines the vertex matrices defined in Eq.~(\ref{eqn:vertexintegral}), recall Section~4.3.
Applying the linear operator $\mathcal{S}(\omega)(\bullet)=(\matrix{J}-i\omega)(\bullet)+(\bullet)\matrix{J}^T$ to the above equation, we find
\begin{align}
\mathcal{S}(\omega) \vec{\matrix{{V}}}(\omega) = \int_0^\infty {\dd \tau} (\partial_\tau {\rm e}^{(\matrix{J}-i\omega) \tau}) \matrix{M} {\rm e}^{\matrix{J}^T \tau} +\int_0^\infty {\dd \tau}  {\rm e}^{(\matrix{J}-i\omega) \tau} \matrix{M} (\partial_\tau {\rm e}^{\matrix{J}^T \tau}).
\end{align}
{By partial integration, we then} obtain
\begin{align}
 \mathcal{S}(\omega) \vec{\matrix{{V}}}(\omega)+\matrix{M}(\omega)=0.
\end{align}
As a special case, we consider $\omega=0$, from which it follows that the integral
\begin{align}
  \matrix{\sigma}=\int_0^\infty {\dd \tau} {\rm e}^{\matrix{J} \tau} \matrix{BB}^T {\rm e}^{\matrix{J}^T \tau}
\end{align}
is a general solution to the Lyapunov equation
\begin{align}
 \matrix{J}\matrix{\sigma}+\matrix{\sigma}\matrix{J}^T+\matrix{BB}^T =0.
\end{align}

\subsection{{Explicit evaluation of vertex functions}}
\label{app:vertexfunctions}

The result of the previous subsection can be applied to {evaluate the vertex functions defined in Section~\ref{sec:loopdiagrams},} as follows. Considering first
\begin{align}
\left[\vec{\matrix{V}}^{j}_{(3)}(\omega)\right]_{\beta\gamma} &\equiv \frac{2\times2}{2!}\qquad 
\begin{tabular}[c]{c}
\begin{fmffile}{diagrams/D6vertex}
\begin{fmfgraph*}(40,40)
\fmfright{i1}
\fmfleft{o1,o2}
\fmf{fermion,lab=$\omega$,lab.dist=-0.3w}{i1,v}
\fmf{boson,label=$\ell$}{o1,v}
\fmf{fermion,label=$\omega+\ell$}{v,o2}
\fmfdot{v}
\fmfv{lab=$j$}{i1}
\fmfv{lab=$\gamma$,lab.d=0.02w}{o1}
\fmfv{lab=$\beta$,lab.d=0.02w}{o2}
\fmfv{lab=$J_{\delta\sigma}^\rho$,lab.dist=-0.75w,lab.a=0}{v}
\end{fmfgraph*}
\end{fmffile}
\end{tabular}
\notag\\
&=
2\int_{-\infty}^{\infty} \frac{\dd \ell}{2\pi}
F_{\beta\delta}(\omega+\ell)
\left(J_{\delta\sigma}^{\rho}\,F_{\sigma j}^\dagger(\omega)\right)
\Delta_{\rho\gamma}(\ell) 
\end{align}
and setting $[\matrix{M}^j_{(3)}]_{\delta\rho}=2\, J_{\delta\sigma}^{\rho}\,F_{\sigma j}^\dagger(\omega)$, we find
\begin{align}
 \vec{\matrix{V}}_{(3)}^j(\omega)
&=\int_{-\infty}^{\infty} \frac{\dd \ell}{2\pi}\matrix{F}(\omega+\ell)\matrix{M}_{(3)}^j(\omega)\matrix{\Delta}(\ell)
\notag\\
&=\int_{-\infty}^{\infty} \dd \tau e^{-i\omega\tau} \matrix{F}(\tau)\matrix{M}_{(3)}^j(\omega)\matrix{\Delta}^T(\tau)
\notag\\
&=\int_{0}^{\infty} \dd \tau e^{(\matrix{J}-i\omega)\tau} \matrix{M}_{(3)}^j(\omega)\matrix{\sigma} e^{\matrix{J}^T\tau},
\end{align}
thus, 
\begin{align}
 \mathcal{S}(\omega)\vec{\matrix{V}}_{(3)}^j(\omega)+\matrix{M}_{(3)}^j(\omega)\matrix{\sigma}=0.
\end{align}
Next, consider the vertex
\begin{align}
\left[\vec{\matrix{V}}^{j}_{(\ast)}(\omega)\right]_{\beta\gamma} &\equiv \frac{1}{2!}\qquad 
\begin{tabular}[c]{c}
\begin{fmffile}{diagrams/D3vertex}
\begin{fmfgraph*}(40,40)
\fmfright{i1}
\fmfleft{o1,o2}
\fmf{fermion,lab=$\omega$,lab.dist=-0.3w}{i1,v}
\fmf{boson,label=$\ell$}{o1,v}
\fmf{boson,label=$\omega+\ell$}{v,o2}
\fmfdot{v}
\fmfv{lab=$j$}{i1}
\fmfv{lab=$\gamma$,lab.d=0.02w}{o1}
\fmfv{lab=$\beta$,lab.d=0.02w}{o2}
\fmfv{lab=$J_{\sigma}^{\rho\delta}$,lab.dist=-0.75w,lab.a=0}{v}
\end{fmfgraph*}
\end{fmffile}
\end{tabular}
\notag\\
&=\frac{1}{2}
\int_{-\infty}^{\infty} \frac{\dd \ell}{2\pi}
\Delta_{\beta\rho}(\omega+\ell)
 \left( J_\sigma^{\rho\delta}F^\dagger_{\sigma j}(\omega)\right)
\Delta_{\delta\gamma}(\ell).
\end{align}
This can be written as the sum of two contributions, letting $[\matrix{M}^j_{(5)}]_{\delta\rho}=(1/2)J_\sigma^{\rho\delta}F^\dagger_{\sigma j}(\omega)$, we have
\begin{align}
 \vec{\matrix{V}}_{(\ast)}^j(\omega)
&=\int_{-\infty}^{\infty} \frac{\dd \ell}{2\pi}\matrix{\Delta}(\omega+\ell)\matrix{M}_{(5)}^j(\omega)\matrix{\Delta}(\ell)
\notag\\
&=\int_{-\infty}^{\infty} \dd \tau e^{-i\omega\tau} \matrix{\Delta}(\tau)\matrix{M}_{(5)}^j(\omega)\matrix{\Delta}^T(\tau)
\notag\\
&=\int_{0}^{\infty} \dd \tau e^{(\matrix{J}-i\omega)\tau}\matrix{\sigma} \matrix{M}_{(5)}^j(\omega)\matrix{\sigma} e^{\matrix{J}^T\tau}
\notag\\
&\qquad+\matrix{\sigma}\left(\int_{0}^{\infty} \dd \tau e^{(\matrix{J}-i\omega)^\dagger\tau} \matrix{M}_{(5)}^j(\omega) e^{\matrix{J}\tau}\right)\matrix{\sigma}
\notag\\
&\equiv \vec{\matrix{V}}_{(5)}^j(\omega)+\matrix{\sigma} \vec{\matrix{V}}_{(6)}^j(\omega)\matrix{\sigma},
\end{align}
thus, the quantities $\vec{\matrix{V}}_{(5)}^j$ and $\vec{\matrix{V}}_{(6)}^j$ satisfy
\begin{align}
 &\mathcal{S}(\omega)\vec{\matrix{V}}_{(5)}^j(\omega)+ \matrix{\sigma}\matrix{M}_{(5)}^j(\omega)\matrix{\sigma}=0,
 \notag\\
 &\vec{\matrix{V}}_{(6)}^j(\omega)\mathcal{S}^\dagger(\omega)+ \matrix{M}_{(5)}^j(\omega)=0.
\end{align}

\subsection{Van Kampen's example}
\label{app:VKex}
The first example in which higher order corrections were probed using the system size expansion was given by van Kampen \cite{vanKampen1976}. He considered the creation and annihilation of electron-hole pairs in a semiconductor following the kinetic scheme
\begin{align}
\varnothing \xrightarrow{k_\text{in}} m X , \ \ X \xrightarrow{X} \varnothing
\end{align}
in the case where $m=1$.
The stoichiometry of these reactions and their associated propensities are given by $\matrix{R}=(m,-1)$ and $\vec{\hat{f}}=(k_\text{in},\phi^2)^T$, respectively. The LNA yields
\begin{align}
 \Delta(\omega)=\frac{2 (2+m) \phi ^2}{16 \phi ^2+\omega ^2},
\end{align}
while the correction to the mean concentration predicted by the macroscopic rate equations is given by
\begin{align}
 \epsilon=-\frac{1}{\Omega^{1/2}}\frac{m+1}{8}.
\end{align}
The intrinsic noise power spectrum can then be expressed using Eq.~(\ref{single:Spectrum}) together with Eqs.~(\ref{single:Delta}), (\ref{single:TSum}) and (\ref{single:QSum}):
\begin{align}
 S(\omega)=\frac{1}{\Omega}\Delta(\omega)+\frac{1}{\Omega^2}\left[
\frac{4 (1+m) \phi ^3 \left((1+m) \phi ^2-(2 m-1) \omega ^2\right)}{\left(4 \phi ^2+\omega ^2\right)^2 \left(16 \phi ^2+\omega ^2\right) }\right] + O(\Omega^{-3}).
\end{align}
The particular case of $m=1$ has been studied by van Kampen \cite{vanKampen}, and has later been corrected by Calisto and Tirapegui \cite{calisto1993}. The Fourier transform of the $\Omega^{-2}$ term in the square brackets yields the correction to the autocorrelation function
\begin{align}
\frac{1}{\Omega} \frac{\phi}{4} {\rm e}^{-2\phi\tau}+  \frac{1}{\Omega^2}
\frac{{\rm e}^{-4 \tau  \phi }+{\rm e}^{-2 \tau  \phi } (2 \tau  \phi -1)}{8}+O(\Omega^{-3}),
\end{align}
which agrees with the result given in \cite{calisto1993}.

\bibliographystyle{iopart-num}
\bibliography{lit}

\providecommand{\newblock}{}
\begin{thebibliography}{10}
\expandafter\ifx\csname url\endcsname\relax
  \def\url#1{{\tt #1}}\fi
\expandafter\ifx\csname urlprefix\endcsname\relax\def\urlprefix{URL }\fi
\providecommand{\eprint}[2][]{\url{#2}}

\bibitem{mcadams1997}
McAdams H and Arkin A 1997 {\em Proc Natl Acad Sci USA\/} {\bf 94} 814--819

\bibitem{gillespie2007}
Gillespie D 2007 {\em Annu Rev Phys Chem\/} {\bf 58} 35--55

\bibitem{ElfEhrenberg}
Elf J and Ehrenberg M 2003 {\em Genome Res\/} {\bf 13} 2475--2484

\bibitem{hayot2004}
Hayot F and Jayaprakash C 2004 {\em Phys Biol\/} {\bf 1} 205

\bibitem{paulsson2005}
Paulsson J 2005 {\em Phys Life Rev\/} {\bf 2} 157--175

\bibitem{vanKampen1976}
Van~Kampen N 1976 {\em Adv Chem Phys\/} {\bf 34} 245--309

\bibitem{grima2010}
Grima R 2010 {\em J Chem Phys\/} {\bf 133} 035101

\bibitem{ramaswamy2012}
Ramaswamy R, Gonz{\'a}lez-Segredo N, Sbalzarini I~F and Grima R 2012 {\em Nat
  Commun\/} {\bf 3} 779

\bibitem{BMC}
Thomas P, Matuschek H and Grima R 2013 {\em BMC Genomics\/} {\bf 14} S5

\bibitem{scott2012}
Scott M 2012 {\em IET Syst Biol\/} {\bf 6} 116--124

\bibitem{chaturvedi1978}
Chaturvedi S and Gardiner C 1978 {\em J Stat Phys\/} {\bf 18} 501--522 ISSN
  0022-4715

\bibitem{thomas2013}
Thomas P, Straube A, Timmer J, Fleck C and Grima R 2013 {\em J Theor Biol\/}
  {\bf 335} 222--234

\bibitem{grima2011}
Grima R, Thomas P and Straube A 2011 {\em J Chem Phys\/} {\bf 135} 084103

\bibitem{grima2011const}
Grima R 2011 {\em Phys Rev E\/} {\bf 84} 056109

\bibitem{srednicki}
Srednicki M 2007 {\em Quantum field theory\/} (Cambridge University Press)

\bibitem{peliti1985}
Peliti L 1985 {\em J Phys (Paris)\/} {\bf 46} 1469--1483

\bibitem{droz1994}
Droz M and McKane A 1994 {\em J Phys A: Math Gen\/} {\bf 27} L467

\bibitem{kamenev2011}
Kamenev A 2011 {\em Field theory of non-equilibrium systems\/} (Cambridge
  University Press)

\bibitem{tauber2014}
T{\"a}uber U 2014 {\em Critical dynamics: a field theory approach to
  equilibrium and non-equilibrium scaling behavior\/} (Cambridge University
  Press)

\bibitem{calisto1993}
Calisto H and Tirapegui E 1993 {\em J Stat Phys\/} {\bf 71} 683--703 ISSN
  0022-4715

\bibitem{warren2006}
Warren P, T{\u{a}}nase-Nicola S and Ten~Wolde P 2006 {\em J Chem Phys\/} {\bf
  125} 144904

\bibitem{ushakov2005}
Ushakov O, W{\"u}nsche H~J, Henneberger F, Khovanov I, Schimansky-Geier L and
  Zaks M 2005 {\em Phys Rev Lett\/} {\bf 95} 123903

\bibitem{mckane2007}
McKane A, Nagy J, Newman T and Stefanini M 2007 {\em J Stat Phys\/} {\bf 128}
  165--191

\bibitem{vanKampen}
van Kampen N 1992 {\em Stochastic processes in physics and chemistry\/} (North
  Holland) ISBN 0444893490

\bibitem{gillespie1992}
Gillespie D 1992 {\em Physica A\/} {\bf 188} 404--425

\bibitem{zinn2002}
Zinn-Justin J 2002 {\em Quantum Field Theory and Critical Phenomena\/}
  (Clarendon Press) ISBN 9780198509233

\bibitem{hanggi1984}
Hanggi P, Grabert H, Talkner P and Thomas H 1984 {\em Phys Rev A\/} {\bf 29}
  371

\bibitem{moloney2006}
Moloney N and Dickman R 2006 {\em Braz J Phys\/} {\bf 36} 1238--1249

\bibitem{plosOne}
Thomas P, Matuschek H and Grima R 2012 {\em PloS one\/} {\bf 7} e38518

\bibitem{gardiner}
Gardiner C 2009 {\em Stochastic methods: a handbook for the natural and social
  sciences\/} (Springer)

\bibitem{keizer1987}
Keizer J 1987 {\em Statistical thermodynamics of nonequilibrium processes\/}
  (Springer)

\bibitem{tauber2012}
T{\"a}uber U 2012 {\em J Phys A: Math Theor\/} {\bf 45} 405002

\bibitem{thomas2011}
Thomas P, Straube A and Grima R 2011 {\em J Chem Phys\/} {\bf 135} 181103

\bibitem{shahrezaei2008}
Shahrezaei V and Swain P 2008 {\em Proc Natl Acad Sci\/} {\bf 105} 17256--17261

\bibitem{prigogine1968}
Prigogine I and Lefever R 1968 {\em J Chem Phys\/} {\bf 48}

\bibitem{westermark2009}
Westermark P, Welsh D, Okamura H and Herzel H 2009 {\em PLoS Comput Biol\/}
  {\bf 5} e1000580

\bibitem{toner2013}
Toner D and Grima R 2013 {\em J Chem Phys\/} {\bf 138} 055101

\bibitem{cianci2012}
Cianci C, Di~Patti F, Fanelli D and Barletti L 2012 {\em Eur Phys J Spec Top\/}
  {\bf 212} 5--22

\bibitem{orszag2008}
Orszag M 2008 {\em Quantum optics: including noise reduction, trapped ions,
  quantum trajectories, and decoherence\/} (Springer Verlag)

\end{thebibliography}

\end{document}